\documentclass[a4paper,twoside]{article}
\newcommand{\affil}[1]{$^{\rm #1}$}
\def\degr{\hbox{$^\circ$}}
\def\arcmin{\hbox{$^\prime$}}
\def\arcsec{\hbox{$^{\prime\prime}$}}
\def\utw{\smash{\rlap{\lower5pt\hbox{$\sim$}}}}
\def\udtw{\smash{\rlap{\lower6pt\hbox{$\approx$}}}}

\def\apj{ApJ}%
\def\apjs{ApJS}%
\def\aap{A\&A}%
\def\mnras{MNRAS}%
\def\pasj{PASJ}%
\usepackage[authoryear]{natbib}
\bibpunct{(}{)}{;}{a}{}{,}
\usepackage{graphicx}
\usepackage{color}
\usepackage{colortbl}
\date{} 
\newcommand{\kms}{\mbox{km\,s$^{-1}$}}
\title{\large\bf\flushleft Characterisation of the Mopra Radio Telescope at 16--50 GHz}
\author{\parbox{\textwidth}{\flushleft
\vspace{-0.5cm}
{\it J.\,S.\,Urquhart\affil{A,E}}, \it M.\,G.\,Hoare\affil{B}, \it C.\,R.\,Purcell\affil{C}, \it K.\,J.\,Brooks\affil{A}, \it  M.\,A.\,Voronkov\affil{A},  \it B.\,T.\,Indermuehle\affil{A}, \it M.\,G.\,Burton\affil{D}, \it N.\,F.\,H.\,Tothill\affil{D} and \it P.\,G.\,Edwards\affil{A}\\
\vspace{0.4cm}
{\small \affil{A}\,Australia Telescope National Facility, CSIRO Astronomy and Space Science, Sydney, NSW 2052, Australia}\\
{\small \affil{B}\,School of Physics and Astronomy, University of Leeds, Leeds, LS2\,9JT, UK}\\
{\small \affil{C}\, 
University of Manchester, Jodrell Bank Centre for Astrophysics, Manchester, M13 9LP, UK}\\
{\small \affil{D}\, 
School of Physics, University of New South Wales, Sydney, NSW 2052, Australia}\\
{\small \affil{E}\,Email: James.Urquhart@csiro.au (ATNF)}}}
\begin{document}
\twocolumn[
\begin{changemargin}{.8cm}{.5cm}
\begin{minipage}{.9\textwidth}
\vspace{-1cm}
\maketitle
%
%
\small{\bf Abstract:} We present the results of a programme of scanning and mapping observations of astronomical masers and Jupiter designed to characterise the performance of the Mopra Radio Telescope at frequencies between 16--50\,GHz using the 12-mm and 7-mm receivers. We use these observations to determine the telescope beam size, beam shape and overall telescope beam efficiency as a function of frequency. We find that the beam size is well fit by $\lambda$/$D$ over the frequency range with a correlation coefficient of $\sim$90\%. We determine the telescope main beam efficiencies are between $\sim$48--64\%  for the 12-mm receiver and reasonably flat at $\sim$50\% for the 7-mm receiver. Beam maps of strong H$_2$O  (22\,GHz) and SiO masers (43\,GHz) provide a means to examine  the radial beam pattern of the telescope. At both frequencies the radial beam pattern reveals the presence of three components, a central `core', which is well fit by a Gaussian and constitutes the telescopes main beam, and inner and outer error beams. At both frequencies the inner and outer error beams extend out to approximately 2 and 3.4 times the full-width half maximum of the main beam respectively. Sources with angular sizes a factor of two or more larger than the telescope main beam will couple to the main and error beams, and therefore the power contributed by the error beams needs to be considered. From measurements of the radial beam power pattern we estimate the amount of power contained in the inner and outer error beams is of order one-fifth at 22\,GHz rising slightly to one-third at 43\,GHz.


\medskip
\medskip
\end{minipage}
\end{changemargin}
]
\small

\section{Introduction}
The 22-m Mopra radio telescope is one element 
of the Australia Telescope (Frater, Brooks \& Whiteoak 1992).
The Mopra observatory is located near Coonabarabran, 
New South Wales, Australia. The
telescope is located at a longitude of 149\degr\,5\arcmin\,58\arcsec\ and a
latitude of 31\degr\,16\arcmin\,58\arcsec\ and at an
elevation of 866 metres above sea level.\footnote{See
http://www.narrabri.atnf.csiro.au/mopra/ for more details.}

Construction of the wheel-on-track telescope (Cooper et al.\ 1992)
was completed in late 1991, with the first astronomical observations being made as part of a VLBI 
session the following week and reported in Tingay et al.\ (1998). In its first years,
Mopra was used for VLBI and single-dish 
observations at centimetre wavelengths (e.g., Koribalski,
Johnston \& Otrupcek 1994). 
In 1995, the telescope was used together with the 
Parkes 64-m telescope as part of Project Phoenix, which observed 
209 solar-type stars over the frequency range 1.2--3.0\,GHz to search 
for possible narrow-band transmissions from extra-terrestrial life
(Tarter 1997).
The Mopra 20/13\,cm feed and receiver was modified from the ATCA 
design to improve performance over the wider Project Phoenix range
--- changes which later enabled Mopra to support VLBI observations 
during the descent of the Huygens probe through the atmosphere
of Titan (Witasse et al.\ 2006).

A 3-mm SIS receiver was tested on the antenna in late 1994, 
with the feed illuminating the central solid-paneled, 15\,m of the dish,
with astronomical 3\,mm observations beginning the following year
(e.g., Elmouttie, Haynes, \& Jones, 1997; Liang et al.\ 1997).
A 12-mm receiver was constructed for Mopra in 1996, primarily to support 
space VLBI observations with the HALCA satellite as part of the
VLBI Space Observatory Programme (Hirabayashi et al.\ 2000).
The compromised performance of the HALCA 12-mm receiver 
(Kobayashi et al.\ 2000) resulted in Mopra support for the VSOP mission 
being at 1.6 and 4.9\,GHz (e.g., Scott et al. 2004; Dodson et al. 2008),
however the ``VSOP'' 12-mm receiver was used in a number of other 
Long Baseline Array VLBI observations (e.g., Greenhill et al.\ 2003)
until it was replaced with an 
MMIC-based receiver (Gough et al.\ 2004) in 2006.

Under an agreement between ATNF and the University of New South Wales
(UNSW), UNSW funded the replacement of the outer perforated panels of
the dish in 1999 with solid panels, and the receiver optics were
modified to illuminate the full 22\,m. Several rounds of holography
using a 30-GHz satellite beacon were used to reduce the surface RMS
error from 270\,$\mu$m in 1999 to 180\,$\mu$m in 2004. 
In 2004, observing capabilities were enhanced with the addition of
an on-the-fly mapping mode, enabling efficient mapping of larger regions of 
interest. The 3-mm SIS receiver was the main receiver used for single-dish observing
during this period, with the telescope being characterised in the
3-mm band by Ladd et al.\ (2005). 
Following the replacement of the 
3-mm SIS receiver in 2005, and the addition of a 12-mm receiver in 2006 and 
7-mm receiver (Moorey et al.\ 2008) in 2008, Mopra now has a suite of
three receivers for single-dish 
observations, covering the frequency ranges
16--28, 30--50 and 76--117\,GHz.  The receivers are
cryogenically cooled Indium Phosphide (InP) High Electron Mobility
Transistor (HEMT) Monolithic Microwave Integrated Circuits (MMICs)
low-noise amplifiers. The receiver systems do not require any manual
tuning and include noise diodes for system temperature
determination, with the 3\,mm receiver also having a paddle for atmospheric attenuation ($\tau$) corrections.
Physically, the three receivers are mounted in a single dewar
(Moorey et al.\ 2008), with a short turret rotation required to 
change between 12-mm and 3-mm observing frequencies. Changing to and from 7-mm observing requires a 
physical translation of the dewar by about 10\,cm, which takes 
several minutes. 

In 2006, the Mopra Spectrometer
(MOPS)\footnote{MOPS was funded in part through a grant instigated by
the University of New South Wales (UNSW) under the Australian Research
Council Grants scheme for Linkage, Infrastructure, Equipment and
Facilities (LIEF), and in part by CSIRO ATNF.}, 
was installed as the main back-end for single-dish observing.
MOPS is a broadband digital filter bank with four
overlapping sub-bands each with a bandwidth of 2.2\,GHz providing a
total of 8.3\,GHz of continuous bandwidth. The spectrometer can be used
in two observing modes: wideband mode and zoom mode. The wideband mode
provides 8.3\,GHz of continuous coverage with a total of 32,384
spectral channels in each polarisation, providing a frequency resolution of 256\,kHz per
channel and a velocity resolution of 3.4, 1.7 and 0.9\,\kms\ at 22, 44
and 90\,GHz, respectively. The second of these modes provides a total
of 16 zoom windows --- four of which can be placed in each of the four
sub-bands --- each with 137\,GHz of bandwidth and 4,096 spectral
channels providing approximately eight times higher frequency and
velocity resolution than the wideband mode.

Observations at 3\,mm are restricted to the
winter months (May--October) when the atmosphere is generally more
stable and the amount of water vapour in the atmosphere is generally
lower. However, observations at 12 and 7\,mm are less affected by atmospheric
fluctuations and water vapour content than at 3\,mm, and thus, can be scheduled
through most of the year. Therefore the installation of the 12-  and 7-mm
receivers has significantly increased the potential science output of
the telescope (e.g., Urquhart et al.\ 2009, Walsh et al.\ 2007, 2008).

In this paper we will present a combination of continuum
and spectral-line observations of Jupiter and H$_2$O and SiO masers
chosen to characterise the telescope's beam size, shape and efficiency
between 16--50\,GHz. To maintain consistency across all three
millimetre bands we have followed the procedures and analyses used by
Ladd et al.\ (2005) to characterise the 3-mm system.

\section{Observations and Data Reduction}

The observations were made between October and December 2009 and consisted of a combination of two orthogonal cross-scans, beam maps, and position-switched observations. During all of these observations the pointing was checked, using the standard five-point procedure, approximately once an hour with corrections found to be $<$10\arcsec. The focus was set to 16.7\,mm since this position was found to be optimum for both the 7- and 12-mm observations. In the following subsections we will describe the details of the various observational modes used and their respective data reduction procedures. No opacity corrections were applied at 12 and 7\,mm as they are small at these frequencies (of order a few percent at the frequencies and elevations of these observations).

\subsection{Jupiter Observations}

A combination of cross-scans and beam maps were made towards Jupiter in order to calculate the telescope's main beam size, shape and efficiency as a function of frequency. Planets are generally used for estimating telescope characteristics since their sizes and fluxes are fairly well modeled. For these observations we chose Jupiter as it is relatively bright, and compact at the frequencies of interest ($\sim$40\arcsec\ during the time of these observations), and its relative motion due to its orbital velocity is small --- less than a tenth of a beam during the time taken for each observation --- and could therefore be neglected. Continuum observations have the advantage of allowing the beam parameters to be calculated as a function of frequency across the whole 12 and 7\,mm frequency range. These observations  were made using the zoom mode with four windows being placed in each 2\,GHz sub-band; the four windows were subsequently averaged together to provide a total bandwidth of $\sim$500\,MHz, across each sub-band.

The cross-scans were made through the centre of Jupiter in the Right Ascension and Declination directions in order to determine the beam size. The cross-scans consisted of a strip of seventeen points spaced at approximately one quarter beamwidths passing through the centre of the planet. The contribution from the sky and telescope was estimated and removed by observing an off-source position at the beginning and end of each scan. To ensure the scans passed through the centre of the planet the telescope pointing was checked immediately before each scan with the pointing model being corrected as necessary.

Beam maps were made of Jupiter to investigate how the telescope beam shape and efficiency change as a function of frequency. To minimise the effects of atmospheric attenuation due to water vapour these mapping observations were made at an elevation of $\sim$70\degr. The map sizes were approximately $9\arcmin \times9\arcmin$ and $6\arcmin\times6\arcmin$ for the 12 and 7\,mm observations respectively. The mapping observations were produced using the on-the-fly mode and consisted of 35 rows scanned in Right Ascension, with each scan separated in Declination by approximately one-quarter beam width ($\sim$30\arcsec\ and $\sim$20\arcsec\ at 12 and 7\,mm respectively), with an off-source reference position being observed before each row. The position of Jupiter was updated at the beginning of each scan to compensate for its orbital motion. The beam maps took $\sim$20 minutes each to complete, with the telescope pointing being checked before and after each map.

The temperature scale for the planet observations is set by the noise calibration system, which must be stable and well-characterised. For the 7-mm band, the noise source is a commercial unit, ELVA-1 model ISSN-22, whereas for the 12-mm band the noise source was built by the ATNF. The electronic drive circuitry necessary to ensure a stable and flat noise output spectrum is internal to the 7-mm band noise source, whereas the 12-mm noise source is driven by an external constant current power supply.  For both bands the units were tested and characterised before being installed and found to be stable, essentially flat-spectrum noise sources.  The noise diode outputs are introduced into the signal path by waveguide couplers, which also make a contribution to the overall performance. The typical performance of the 7-mm couplers (installed both at Mopra and the ATCA) is described by Moorey et al (2008). The 12-mm coupler has comparable characteristics, with some amplitude variability above 25\,GHz.

\subsection{H$_2$O and SiO Maser Beam Maps}

Although planets are the preferred astronomical sources for determining beam efficiencies, their relatively low signal to noise ratio (SNR), and the difficulty removing the sky emission due to atmospheric variations, makes continuum observations less suitable for determining the telescope's beam pattern. A better choice is to make observations of strong astronomical masers which provide high SNR maps of the telescope beam patterns, primarily due to the better baseline subtraction. However, unlike continuum observations of a planet, that can be made at all available frequencies to each receiver, observations of masers can only be made at a few well defined frequencies, and consequently only provide information at these frequencies.

We have therefore made beam maps of the bright H$_2$O and SiO masers (at 22.24 and 43.12\,GHz respectively) associated with Orion SiO. These two masers allow us to investigate the telescope beam pattern in both the 12- and 7-mm bands. These maps are centred on the position of Orion SiO ($\alpha_{J2000}$ = 5:35:14.5 and $\delta_{J2000}$ = $-$6:22:29.6) and covered a $20\arcmin\times20\arcmin$ and a $10\arcmin\times10\arcmin$ region at 22.24 and 43.12\,GHz, respectively. 

Each map consisted of a total of 25 on-source scans, each individual scan was separated by 51\arcsec\ and 26\arcsec\ for the 22.24 and 43.12\,GHz maps, respectively. Each map required approximately one hour to complete. Two maps were produced for each maser frequency using orthogonal scan directions. These were subsequently combined to reduce striping  (which can result from fluctuations in the sky emission during the observations) and improve the signal to noise in the final map.

\subsection{Data Reduction}

\begin{sloppypar}
The cross-scans and position-switched observations were reduced using the ATNF spectral analysis package (ASAP).\footnote{http://www.atnf.csiro.au/computing/software/asap/.} Each cross-scan consisted of seventeen separate measurements taken across the planet. For each point the average antenna temperature measurement is calculated across the zoom bands. The background antenna temperature obtained from the reference position is then subtracted to produce a plot of antenna temperature as a function of angular offset across the planet (see example presented in Figure\,1).
\end{sloppypar}

\begin{sloppypar}
The beam maps were reduced using LIVEDATA and GRIDZILLA packages available from the ATNF.\footnote{Both software packages are available from http://www.atnf.csiro.au/people/mcalabre/livedata.html.} LIVEDATA performs a bandpass calibration for each row using the off-source data followed by fitting a user-specified polynomial to the spectral baseline. GRIDZILLA grids the data to a user-specified weighting and beam parameter input. The data were weighted using the periodic $T_{\rm{sys}}$ measurements and a cell size of 0.25\arcmin\ and 0.15\arcmin\ was used to grid the 12- and 7-mm data, respectively. 
\end{sloppypar}

\begin{sloppypar}

A Gaussian smoothing kernel was used to produce the beam maps with a FWHM and a radial cutoff of 2\arcmin\ for the 12\,mm data and 1\arcmin\ for the 7\,mm data; these values were chosen to minimise smoothing and obtain a resolution in the reduced maps similar to the native resolution of the telescope at the observed frequencies. However, tests show that the actual choice of the full-width at half maximum (FWHM) used for the smoothing does not have a significant affect on the results.

\end{sloppypar}

\begin{figure}
\begin{center}
\includegraphics[width=0.45\textwidth, trim=25 0 25 0]{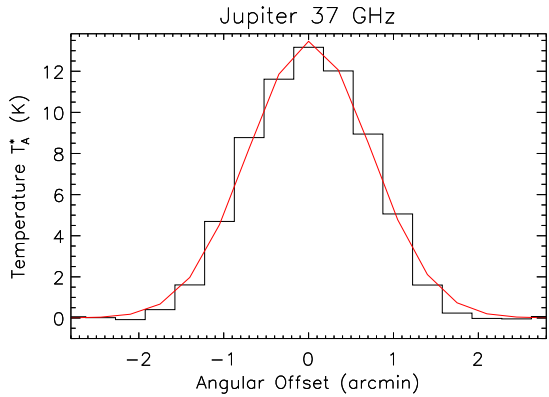}

\caption{Example of the Jupiter cross-scan results. The profile is an average of the two orthogonal scans made at 37\,GHz. The Gaussian fit is shown as a solid red line.}\label{fig:cross_scan_plot}
\end{center}
\end{figure}

\section{Results and Analysis}

\subsection{Telescope Main Beam Size}

\begin{table}
\begin{center}
\caption{Telescope main beam size (FWHM) as a function of frequency.}\label{tbl:cross_scan}
\begin{tabular}{ccc}
\hline Centre Frequency & Beam Size & Beam Error \\
(GHz)& (\arcmin) & (\arcmin) \\
\hline 
17	&	2.68	&	0.52	\\
19	&	2.42	&	0.40	\\
20	&	2.33	&	0.40	\\
21	&	2.25	&	0.38	\\
22	&	2.11	&	0.30	\\
23	&	2.05	&	0.30	\\
24	&	1.99	&	0.23	\\
26	&	1.81	&	0.25	\\
31	&	1.37	&	0.10	\\
33	&	1.30	&	0.08	\\
35	&	1.29	&	0.06	\\
37	&	1.23	&	0.06	\\
39	&	1.19	&	0.05	\\
41	&	1.12	&	0.04	\\
43	&	1.14	&	0.05	\\
45	&	1.08	&	0.04	\\
47	&	1.04	&	0.04	\\
49	&	0.99	&	0.04	\\

\hline
\end{tabular}
\end{center}
\end{table}

To measure the change in beam sizes as a function of frequency the orthogonal cross-scans of Jupiter were averaged together and fitted with a Gaussian profile. In Figure\,\ref{fig:cross_scan_plot} we present an example of the temperature profile obtained across the planet from the cross-scan observations and the Gaussian fit to the data. The FWHM from the fit is a convolution of the actual beam size and the disk of the planet. The angular size of Jupiter during the observations was $\sim$40\arcsec, corresponding to $\sim$20\% of the telescope beam size at 12\,mm and 60\% at 7\,mm.

To calculate the telescope beam size we need to deconvolve the observed FWHM and the planetary disk using:

\begin{eqnarray}
\theta_{\rm{mb}} = \sqrt{\left(\theta_{\rm{o}}^2-\frac{{\rm{ln2}}}{2}\theta_{\rm{planet}}^2\right)},
\end{eqnarray}

\noindent where $\theta_{\rm{planet}}$ is the angular diameter of the planet, $\theta_{\rm{o}}$ the observed and $\theta_{\rm{mb}}$ the deconvolved main beam, FWHM respectively. In Table\,\ref{tbl:cross_scan} we present the deconvolved beam sizes and their associated 1$\sigma$ error derived from the Gaussian fit.

\begin{figure}
\begin{center}
\includegraphics[width=0.45\textwidth, trim=25 0 25 0]{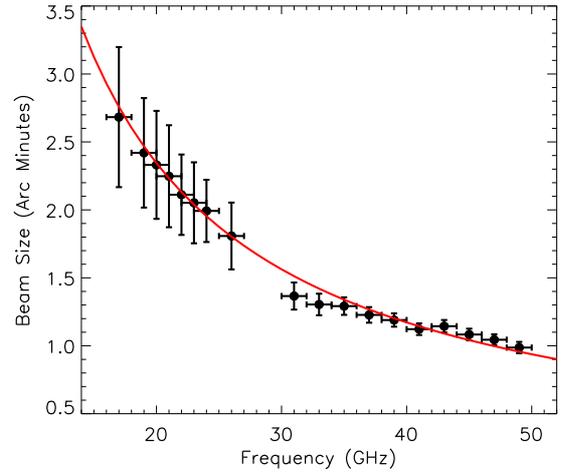}

\caption{Plot showing the dependence of the telescope's FWHM beam size as a function of frequency. The data points are shown as filled circles, the horizontal error bar indicates the bandwidth the data has been taken over while the vertical error bar indicates the 1$\sigma$ error to the fit to the FWHM. The red line shows the theoretical FWHM beam size assuming $\theta=\lambda/D$ (see text for details). }\label{fig:beam_plot}
\end{center}
\end{figure}

In Figure\,\ref{fig:beam_plot} we show the FWHM telescope beam sizes as a function of frequency. In this plot the horizontal and vertical error bars indicate the frequency range the measurement covers and the 1$\sigma$ error the error in the FWHM measurement. The larger fractional errors seen at lower frequencies is related to the increased levels of beam-dilution, resulting in a lower SNR. Additionally, we also plot the theoretical beam sizes (solid red line) calculated assuming $\theta=\lambda/D$, where $\theta$ is measured in radians, $\lambda$ is the observed wavelength and $D$ is the diameter of the telescope. Inspection of the derived and theoretical beams sizes reveals a strong correlation with a correlation coefficient = 0.90.

\subsection{Telescope Beam Pattern}

\begin{figure*}
\begin{center}
\includegraphics[width=0.45\textwidth]{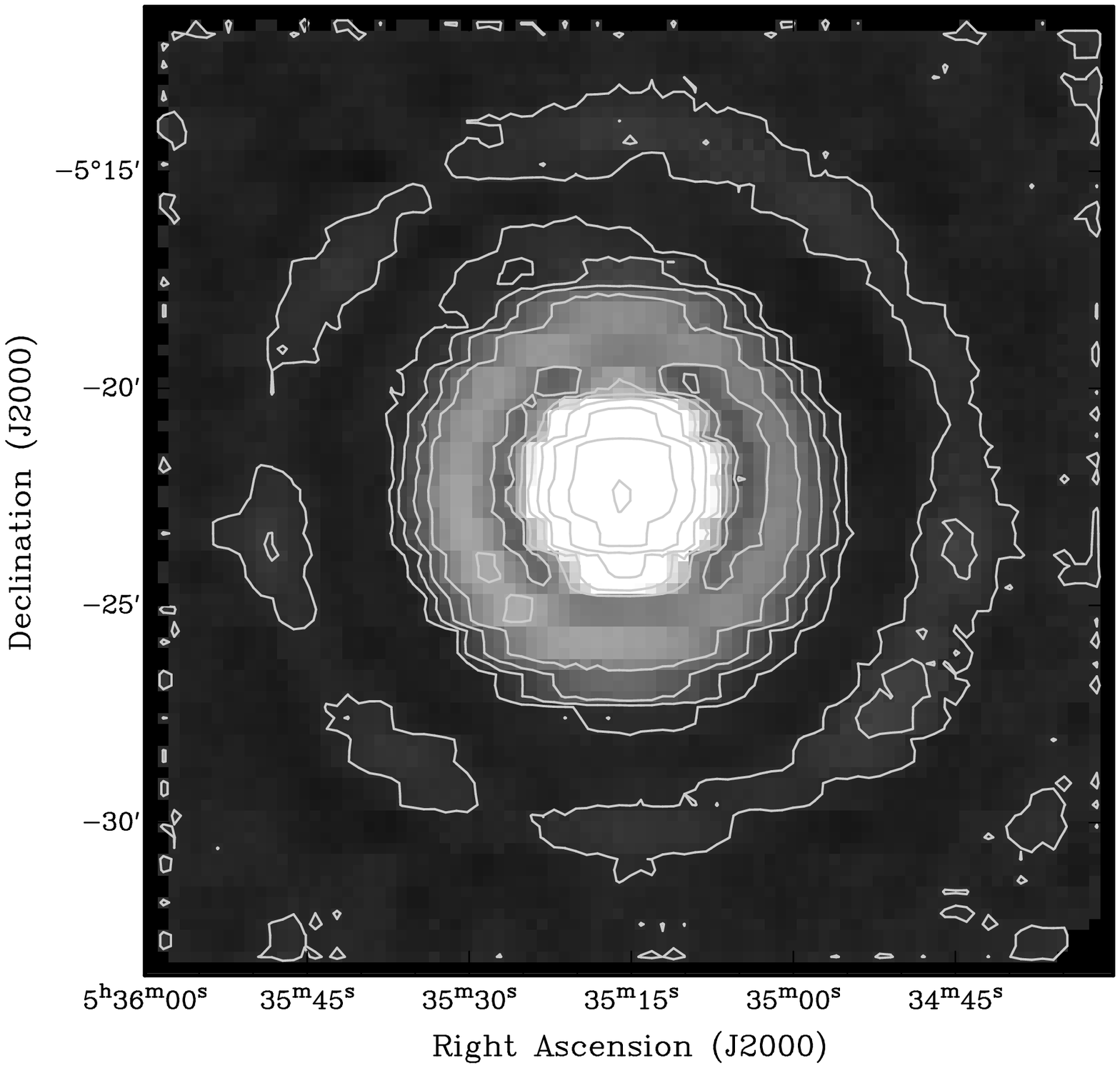}
\includegraphics[width=0.45\textwidth]{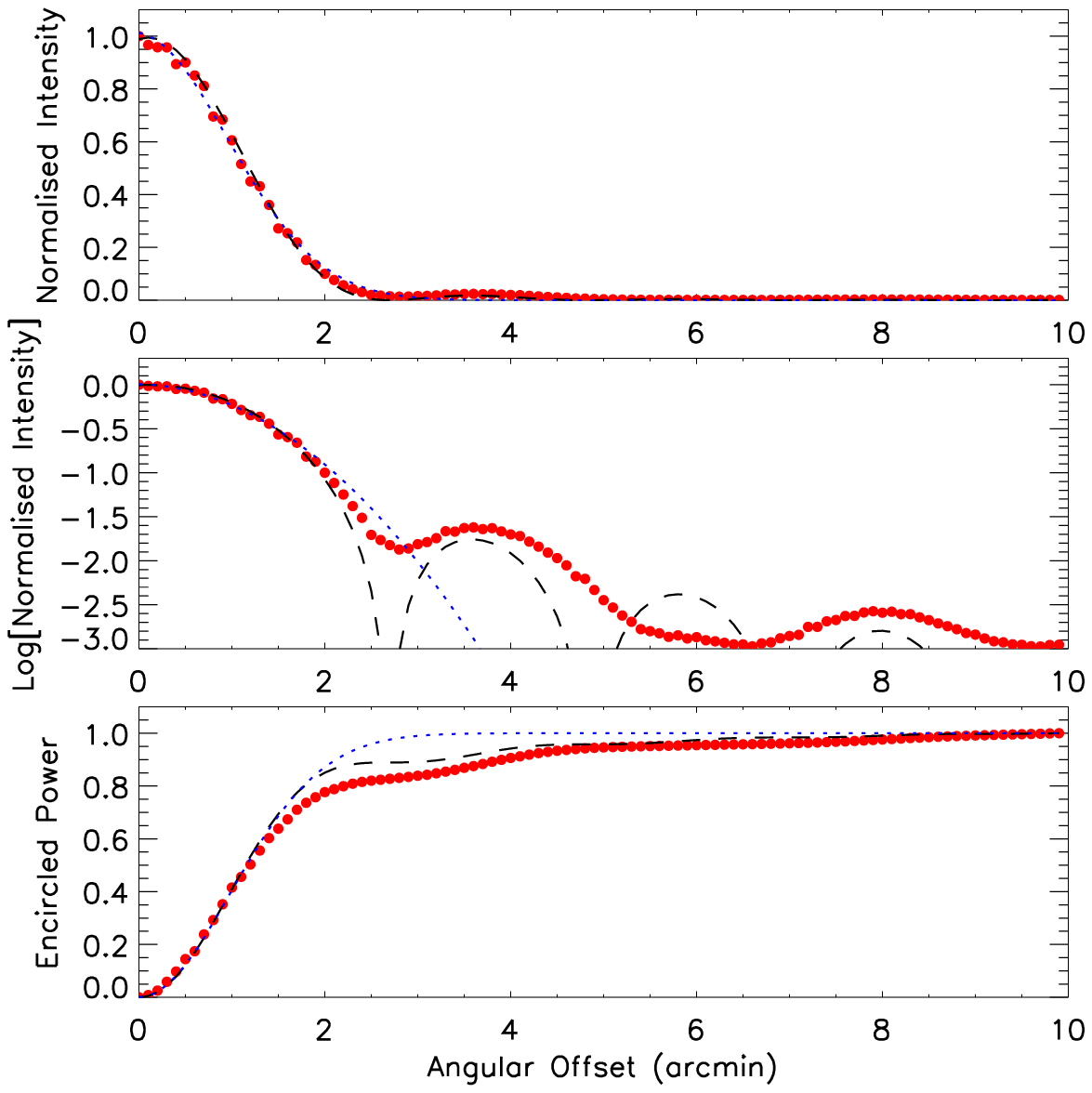}
\vspace{5pt}
\includegraphics[width=0.47\textwidth, trim= 10 0 0 10]{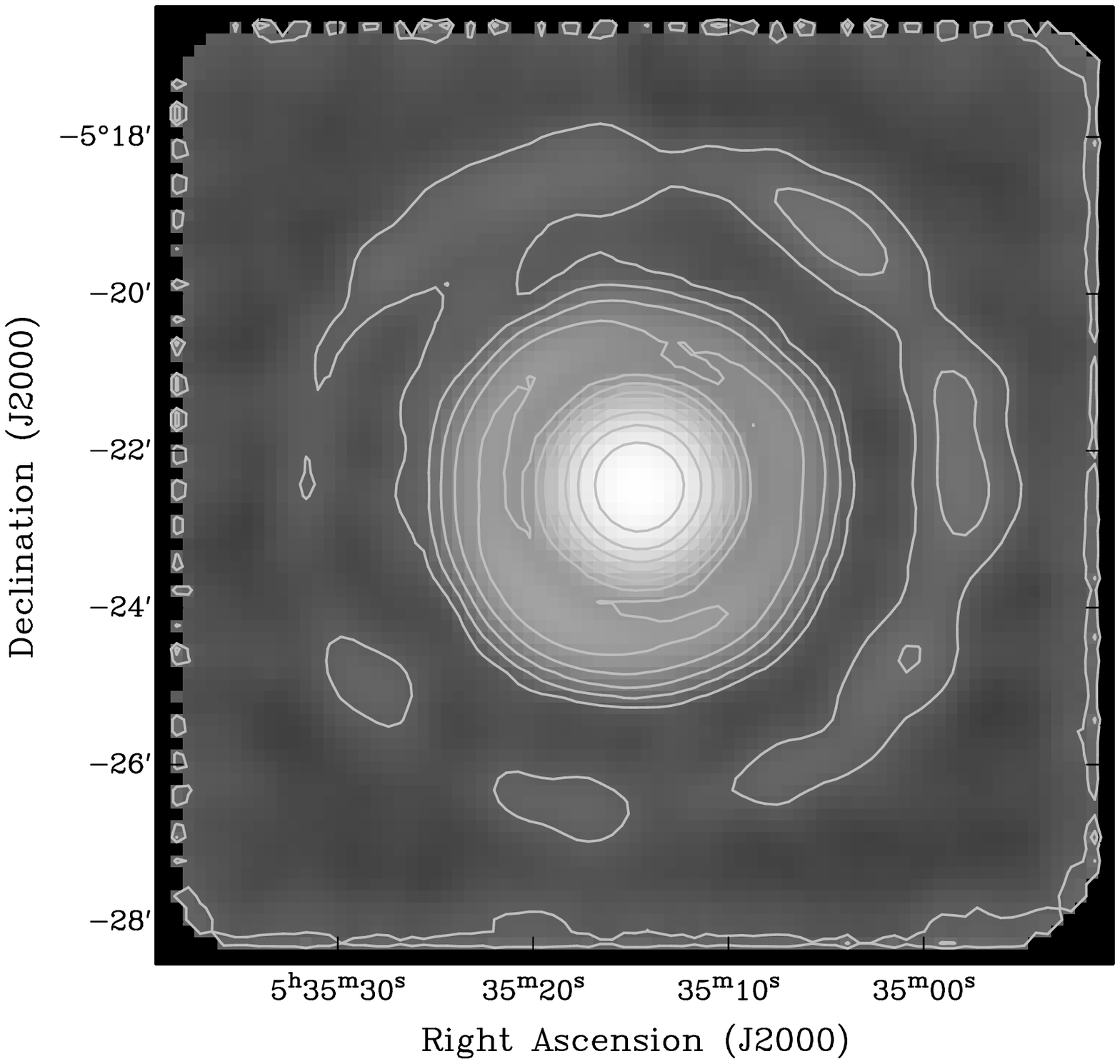}
\includegraphics[width=0.45\textwidth]{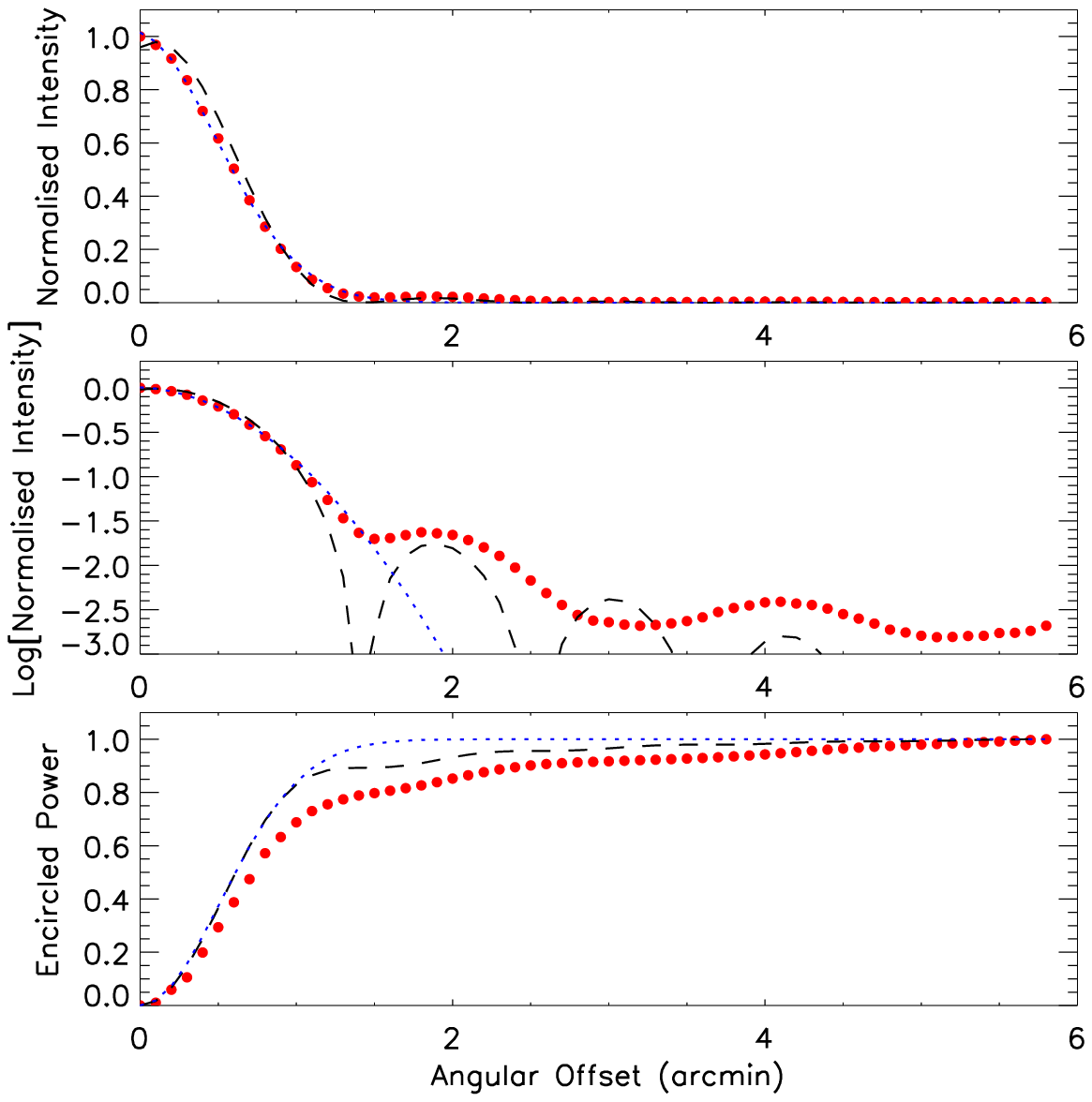}

\caption{Left panels: Integrated 22-GHz (top) and 43-GHz (bottom) maps of the H$_2$O and SiO masers associated with Orion\,KL. The 22- and 43-GHz maps cover a region of $\sim$20\arcmin $\times$ 20\arcmin\ and $\sim$10\arcmin $\times$ 10\arcmin\ respectively. These maps clearly show a central `core' of emission associated with the telescopes main beam and both the inner and outer error beams. Contours have been added to highlight the emission associated with the error beams which have intensities of only a few percent of the main beam and are relatively weak. The first contour starts at 9 and 10\,K\,km\,s$^{-1}$ for the 22- and 43-GHz maps, respectively, and increase in steps determined by a power law of index 1.5.  In the right panels we present azimuthally averaged beam profiles corresponding to the integrated map presented to their left. For both the 22 and 43\,GHz integrated maps we plot the normalised intensity (top), the logarithm of the normalised intensity (middle), and the encircled power (bottom), as a function of radial offset in arcminutes. In addition to the data shown in red, we also plot the Airy profile expected from a fully illuminated 22-m aperture (black dashes) and the results of a Gaussian fit to the central `core' (dotted blue). }\label{fig:beam_pattern}
\end{center}
\end{figure*}

To determine the telescope beam pattern, maps of the 43.12\,GHz SiO and 22.24\,GHz H$_2$O masers associated with Orion KL were produced. In the upper and lower left panels of Figure\,\ref{fig:beam_pattern} we present the integrated emission maps of the H$_2$O and SiO masers respectively. The reference position used for these mapping observations was offset by 1\degr\ in declination from the map centre, and therefore, these maps are sensitive to both the inner and outer error beams. Inspection of these images clearly reveals the telescope's main beam and both inner and outer error beams --- contours have been added to emphasise the emission associated with the outer error beam, which is significantly weaker   than the emission associated with the main beam. In the upper and lower right panels of Figure\,\ref{fig:beam_pattern} we present the plots of the azimuthally averaged $\sim$22\,GHz and $\sim$43\,GHz radial beam profiles respectively (see figure caption for a description of individual plots).

\begin{table}
\begin{center}
\caption{Summary of radial beam pattern derived parameters.}\label{tbl:beam_summary}
\begin{tabular}{clcc}
\hline 
Frequency & Beam & Angular & Encircled \\
& Component&Size & Power\\
(GHz)& &(\arcmin) & (\%)\\
\hline 
22	&	Main	&0.0--2.8	&83	\\
	&	Inner	&2.8--6.5	&13	\\
	&	Outer	&6.5--9.5	&4	\\
43	&	Main	&0.0--1.3	&77	\\
	&	Inner	&1.3--3.2	&15	\\
	&	Outer	&3.2--5.5	&8	\\

\hline
\end{tabular}
\end{center}
\end{table}

\begin{figure*}[!ht]
\begin{center}
\includegraphics[width=0.32\textwidth, trim=50 0 50 0]{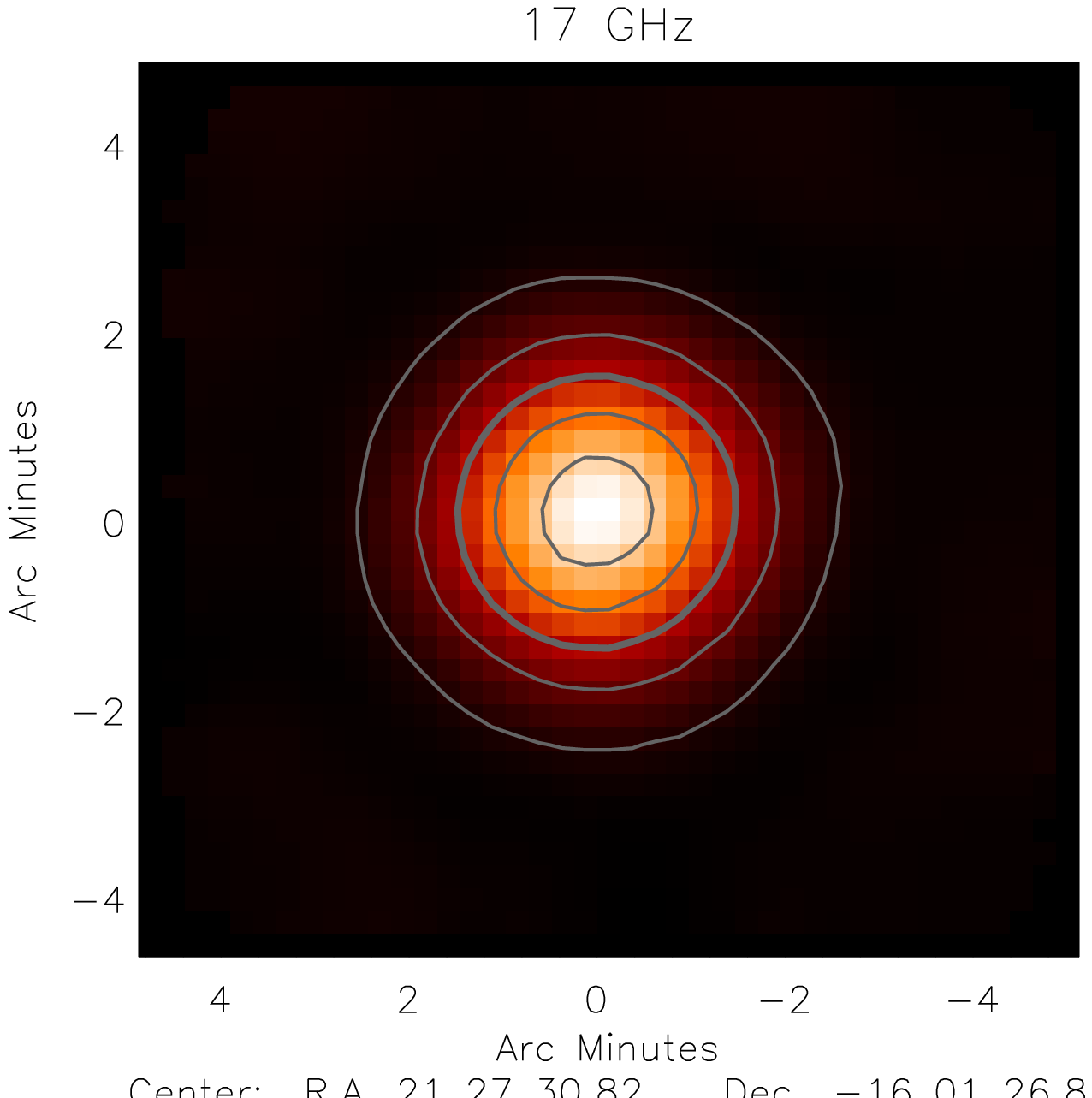}
\includegraphics[width=0.32\textwidth, trim=50 0 50 0]{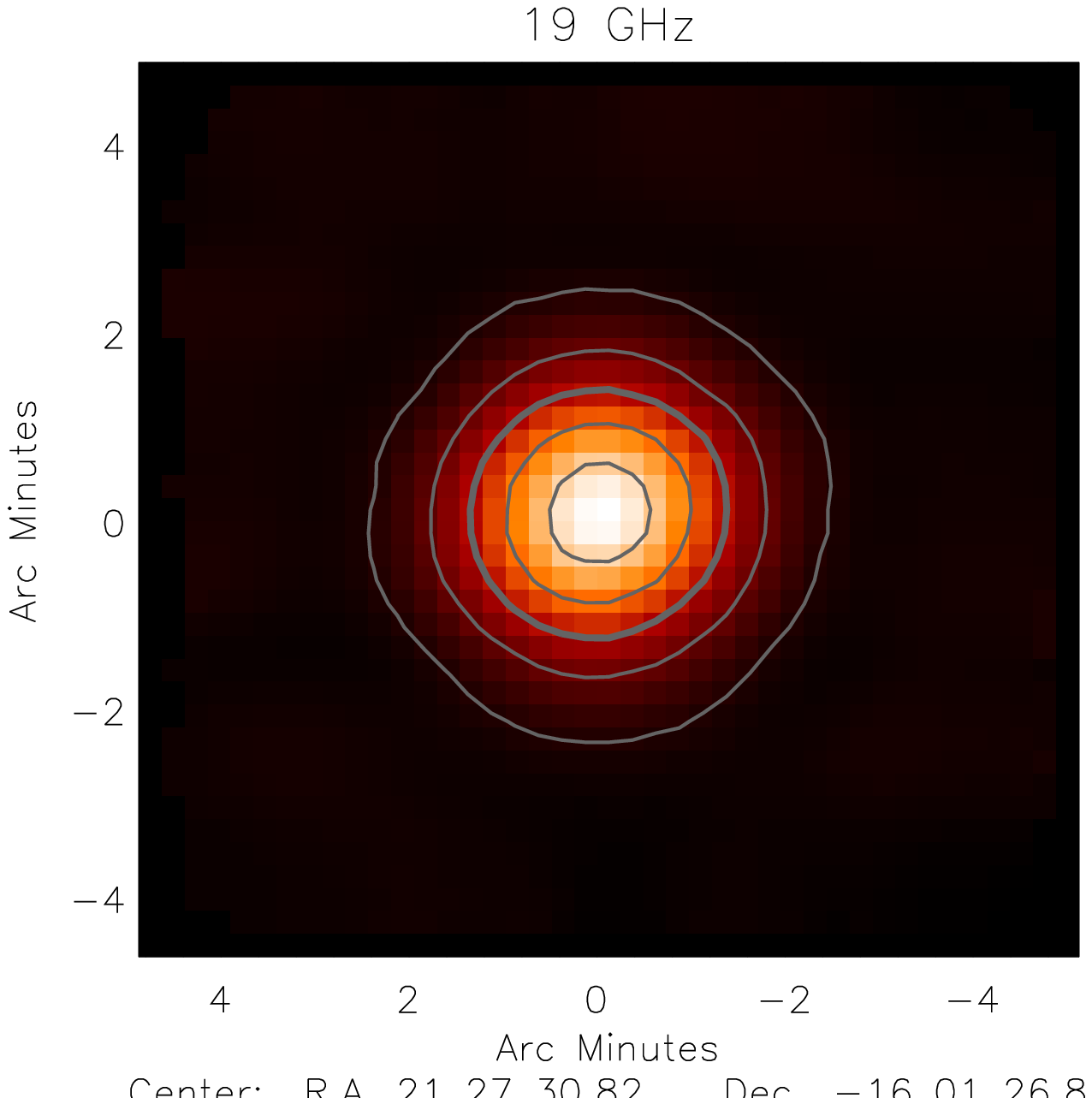}
\includegraphics[width=0.32\textwidth, trim=50 0 50 0]{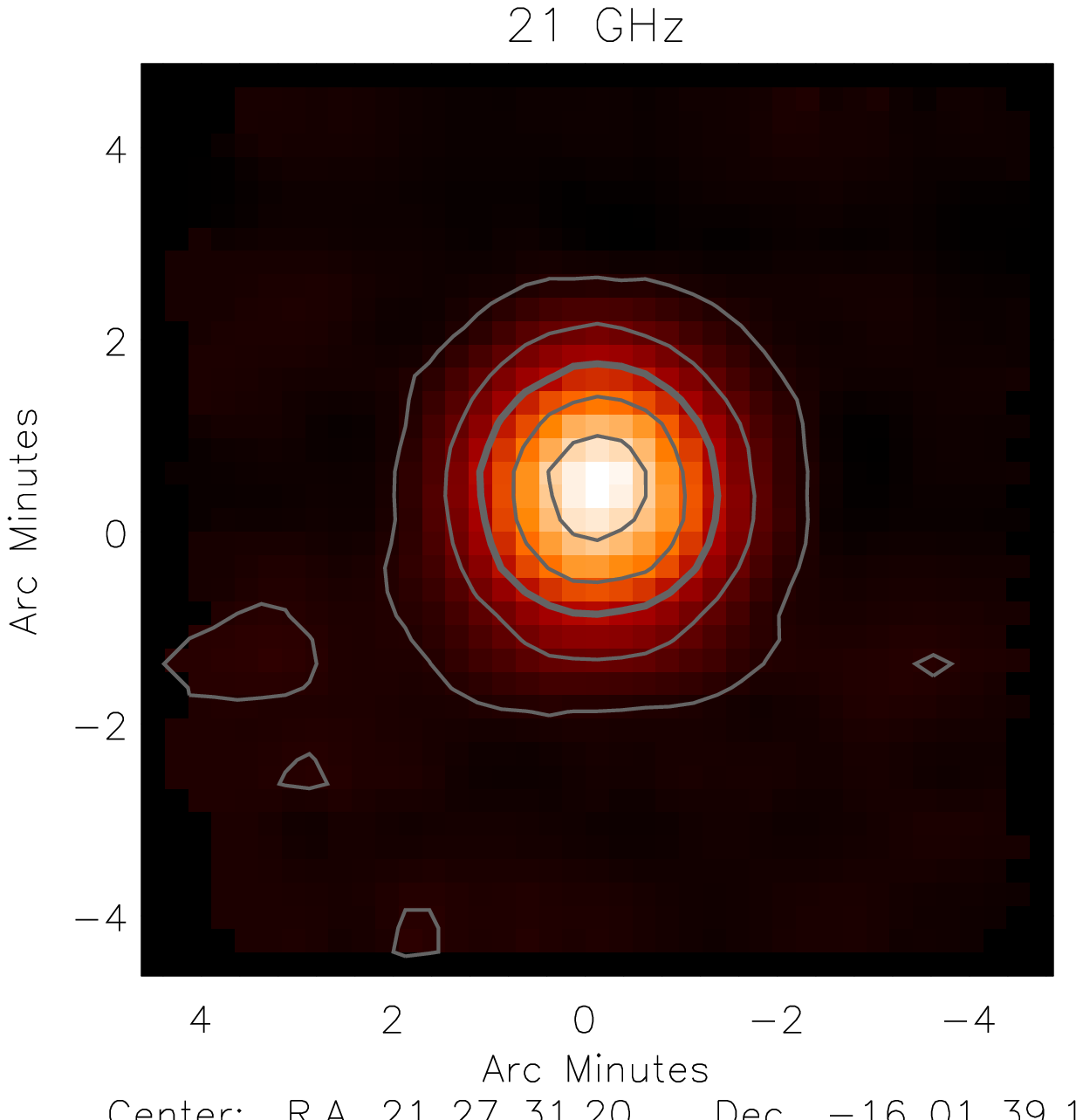}\\
\includegraphics[width=0.32\textwidth, trim=50 0 50 0]{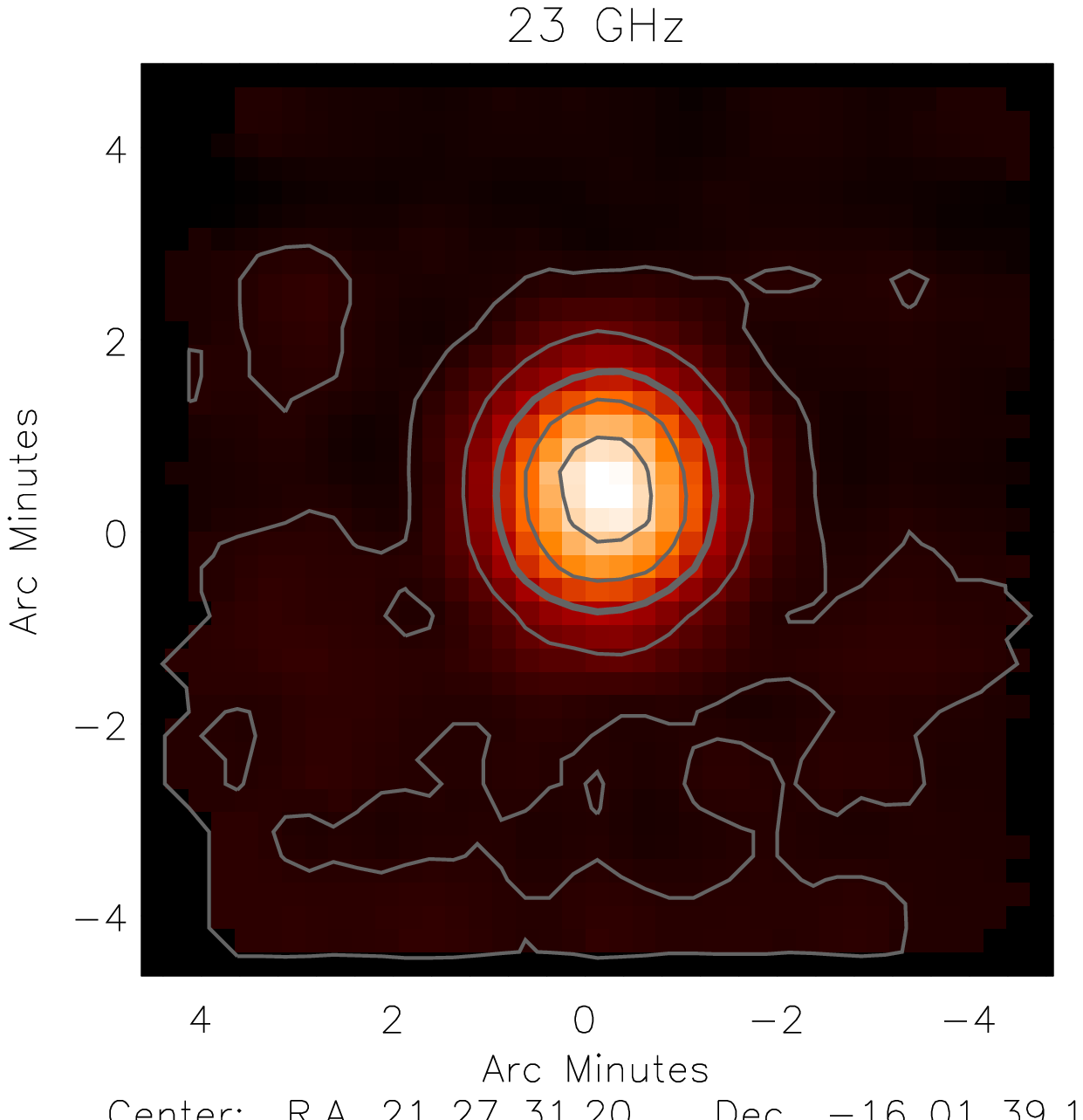}
\includegraphics[width=0.32\textwidth, trim=50 0 50 0]{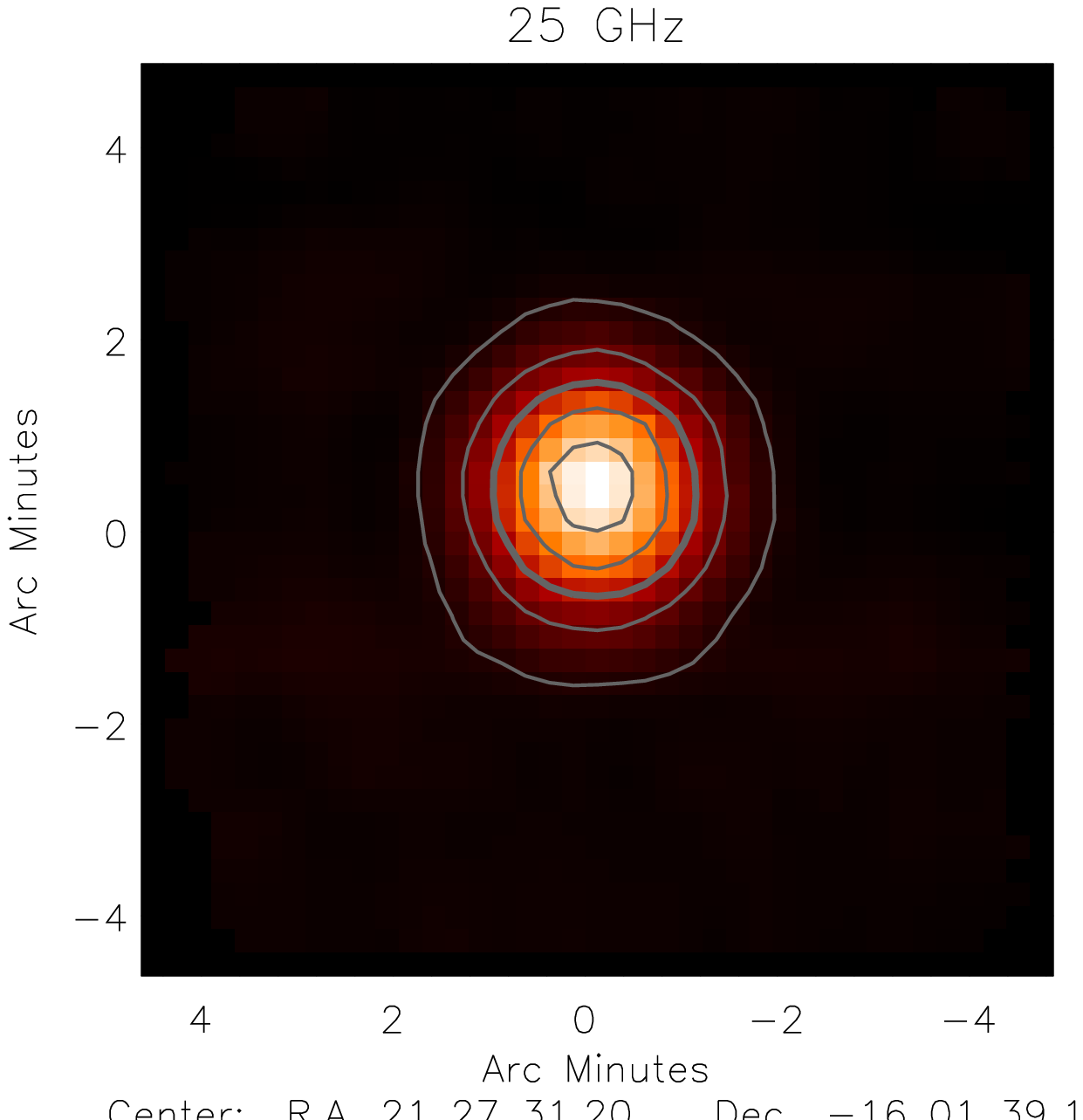}
\includegraphics[width=0.32\textwidth, trim=50 0 50 0]{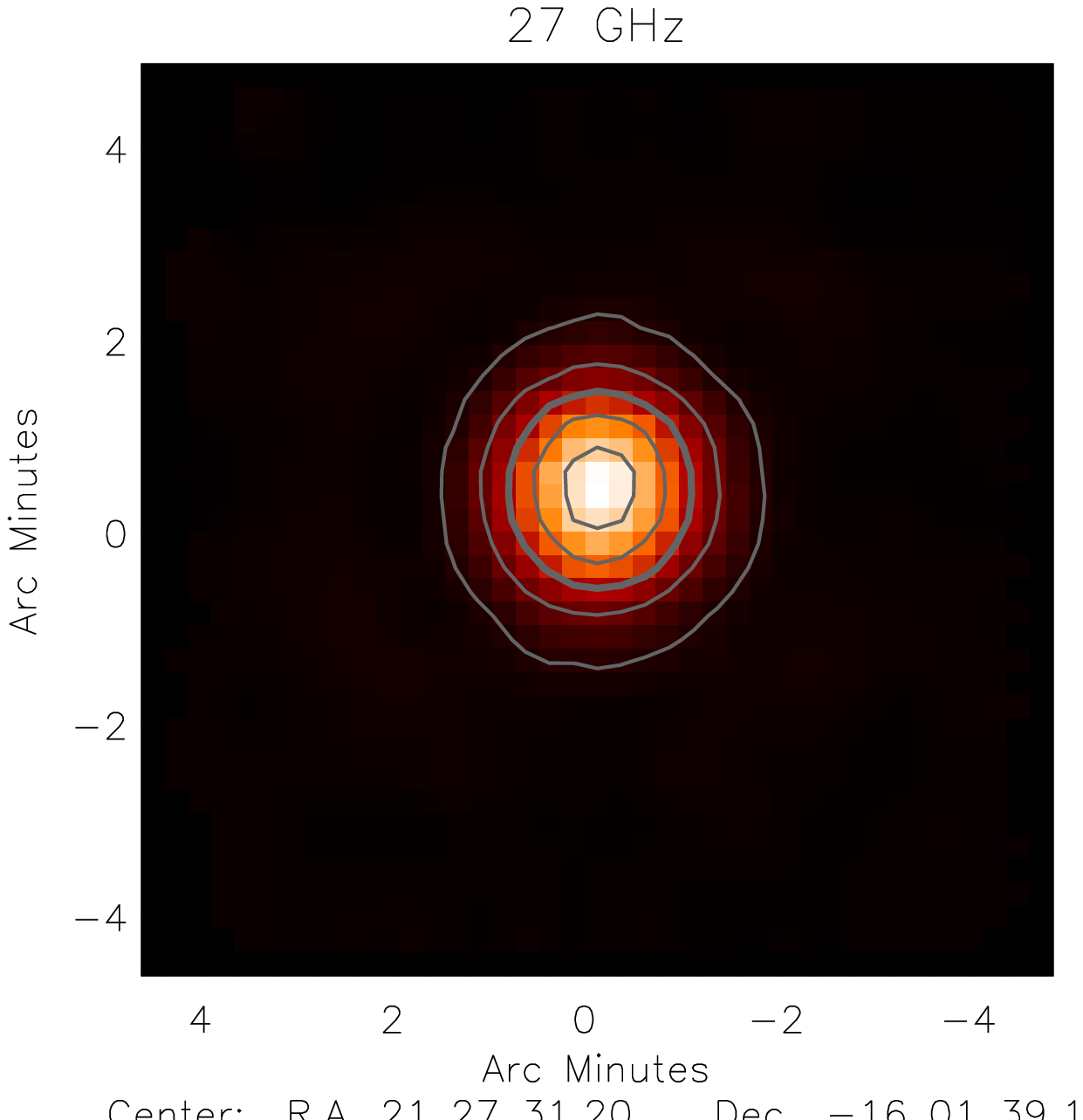}

\caption{Maps of the telescope beam produced from observations of Jupiter using the 12\,mm receiver. Each map covers a 2\,GHz frequency window and is separated by 2\,GHz interval through the 16--28\,GHz frequency range of the 12-mm receiver. Contours have been added to illustrate the shape of the beam at the various frequencies; these start at 10\% of the peak emission and increase in steps of 20\%.}\label{fig:12mm_beam_maps}
\end{center}
\end{figure*}

Inspection of both the H$_2$O and SiO maser integrated emission map shown in Figure\,3 (left panels) reveals a central core and two weaker azimuthal components resulting from the telescope sidelobes. Following Ladd et al.\,(2005) we refer to these as the `main beam', `inner error beam' and `outer error beam'. All three structures can be clearly distinguished and appear to be circular in shape. The azimuthally averaged radial plots presented in the right panels of Figure\,3 of the two maser transitions reveals similar patterns. The central `core' emission in both sets of plots are well-fit with Gaussians of FWHM 2.35\arcmin\ and 1.2\arcmin\ for the H$_2$O and SiO maser respectively. In both cases the main beam sizes determined by the fit to the maser emission is approximately the same as determined using the cross-scan data presented in Section\,3.1. Comparing the expect Airy patterns with the data we see the main beam and, inner and outer error beam for the two masers correspond well with the expected position and intensity ratios of the first, second and fourth components of the Airy pattern. However, there is no counterpart in either sets of data that would correspond with the position of the third Airy component, and it is unclear what might be responsible for the suppression of this particular sidelobe.

Measuring the total encircled power within the main beam at 22\,GHz and
43\,GHz we find values of $\sim$83\% and $\sim$77\% respectively, which
is marginally better than the 75\% reported by Ladd et al. (2005) in
their analysis of the telescope's 3\,mm beam pattern. However, this
means that the error beams can make a significant contribution to the
measured flux for extended sources and need to be taken into account
when estimating the beam efficiency (this will be discussed in more detail in Section 3.3.2). In Table\,\ref{tbl:beam_summary}
we present a summary of the angular extent and the total power
encircled in the main, and inner and outer error beams.

\subsection{Beam Efficiency}

\begin{figure}
\begin{center}
\includegraphics[width=0.23\textwidth, trim=50 0 50 0]{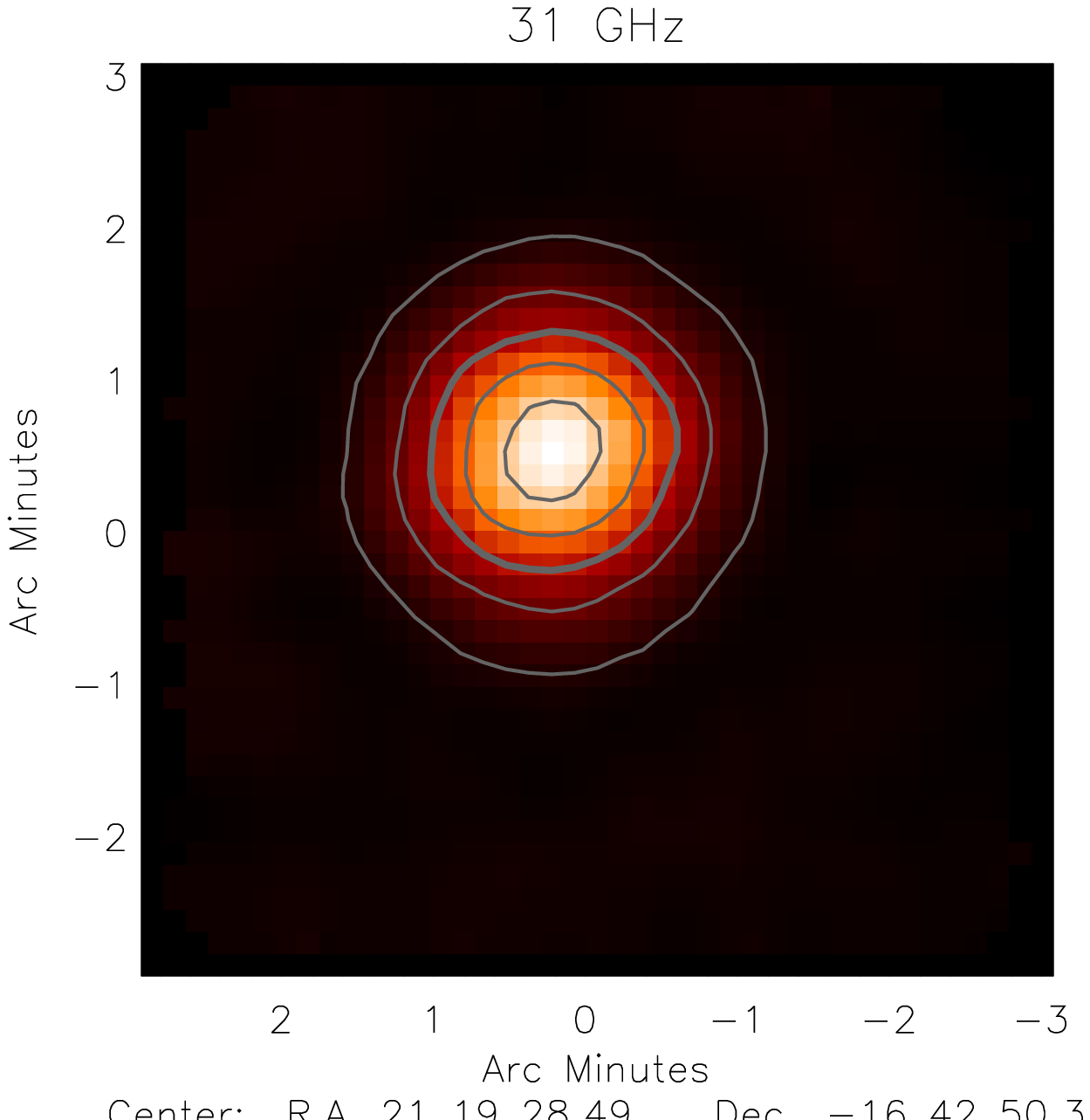}
\includegraphics[width=0.23\textwidth, trim=50 0 50 0]{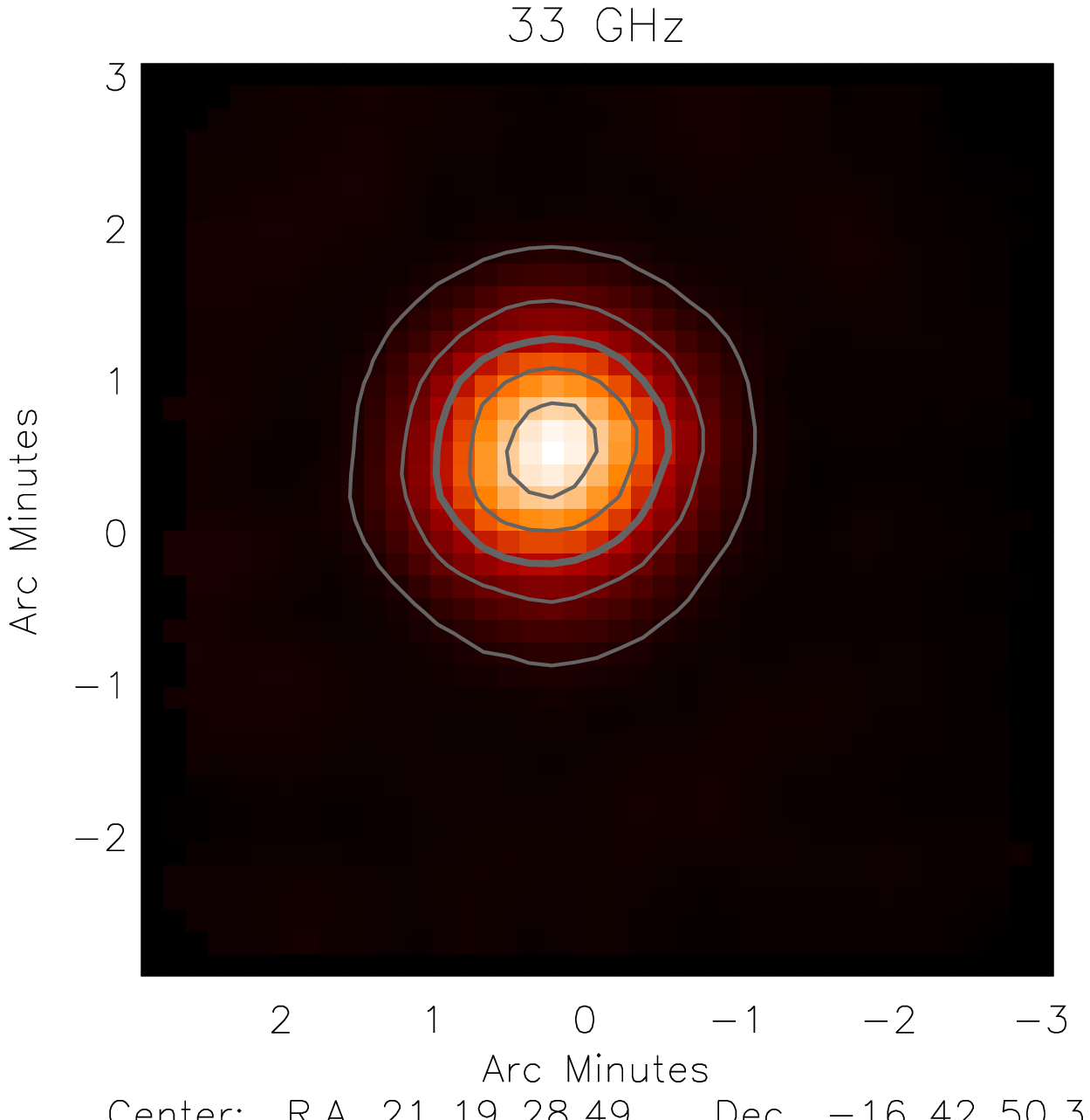}\\
\includegraphics[width=0.23\textwidth, trim=50 0 50 0]{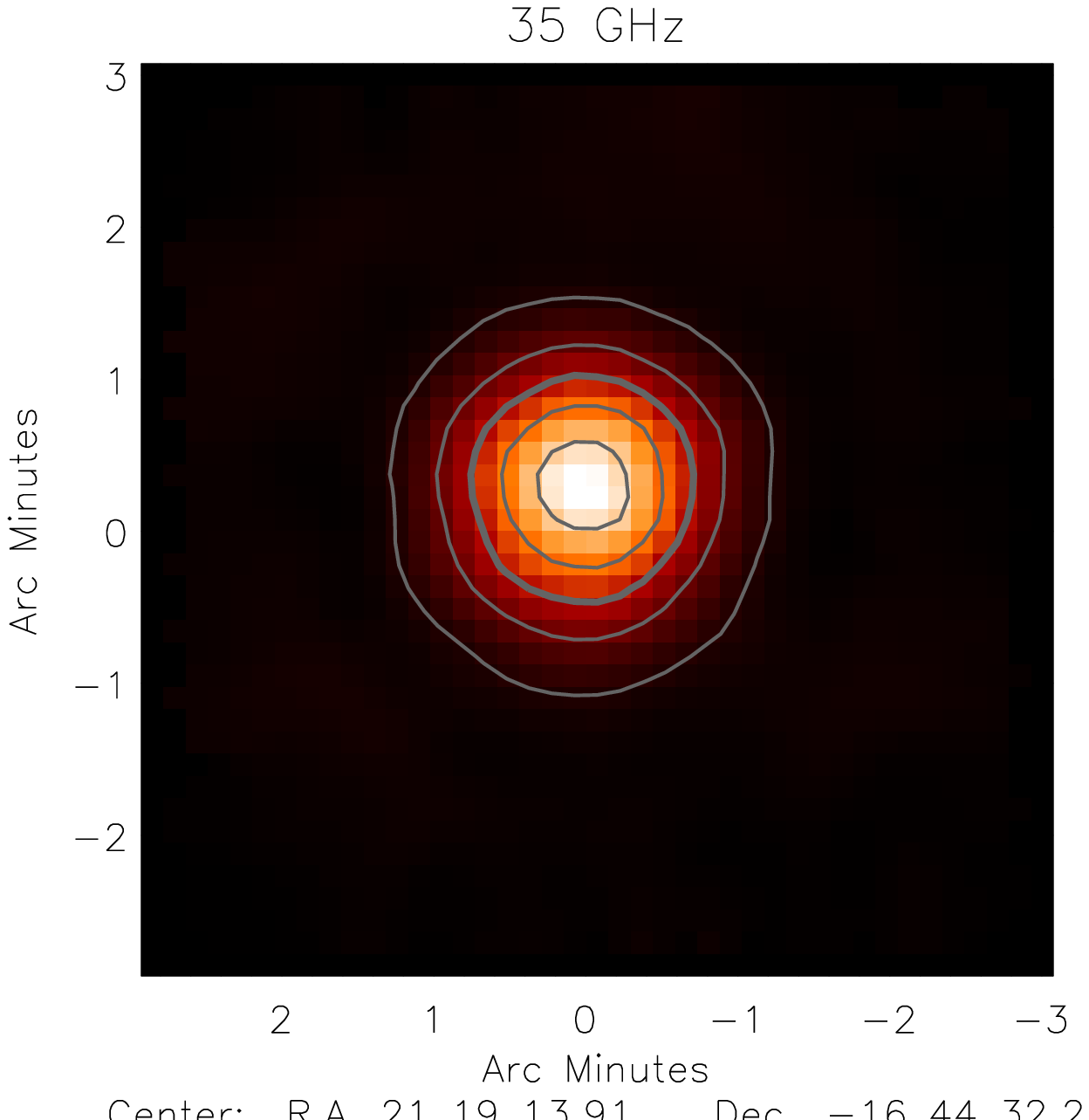}
\includegraphics[width=0.23\textwidth, trim=50 0 50 0]{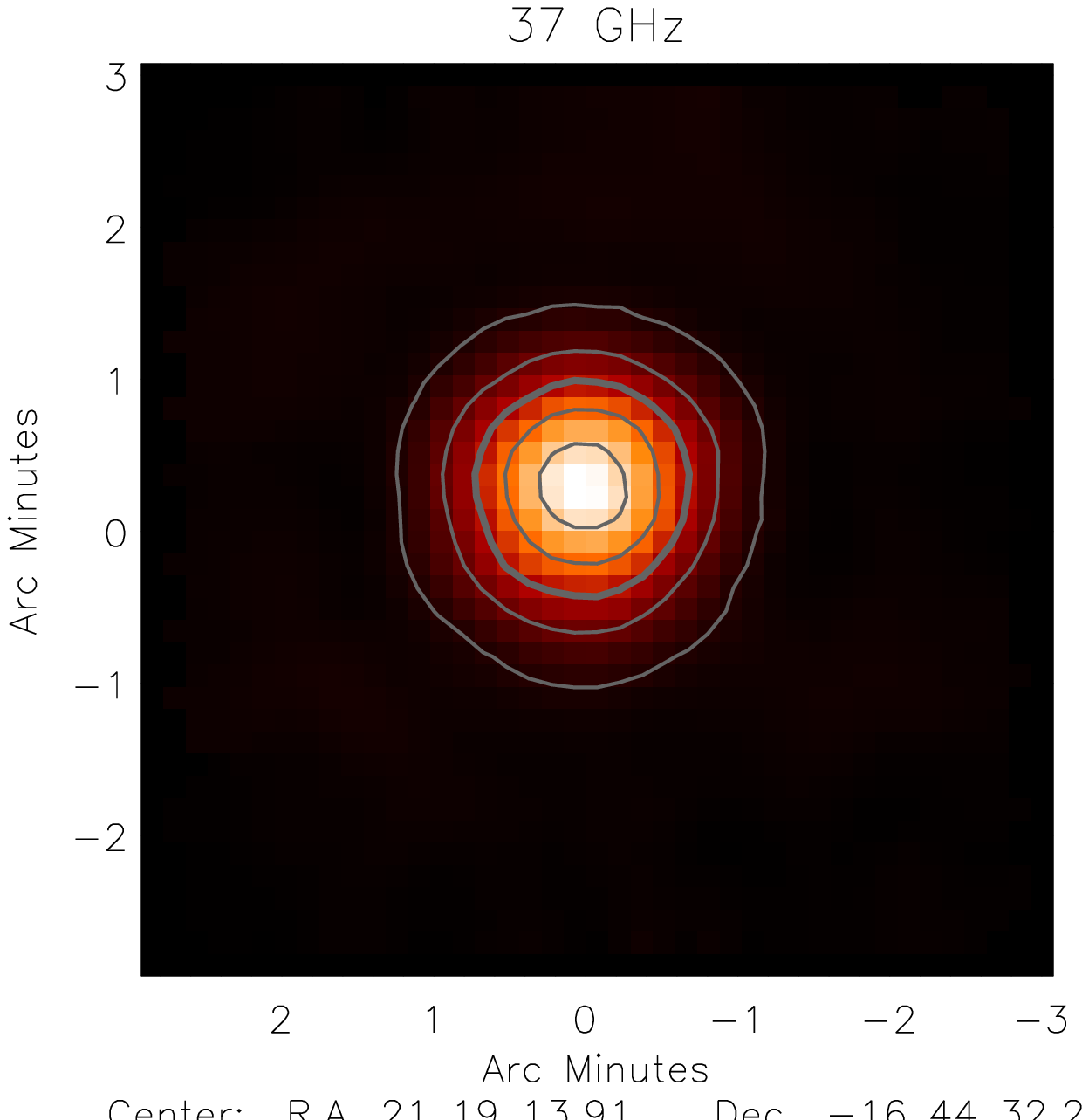}\\
\includegraphics[width=0.23\textwidth, trim=50 0 50 0]{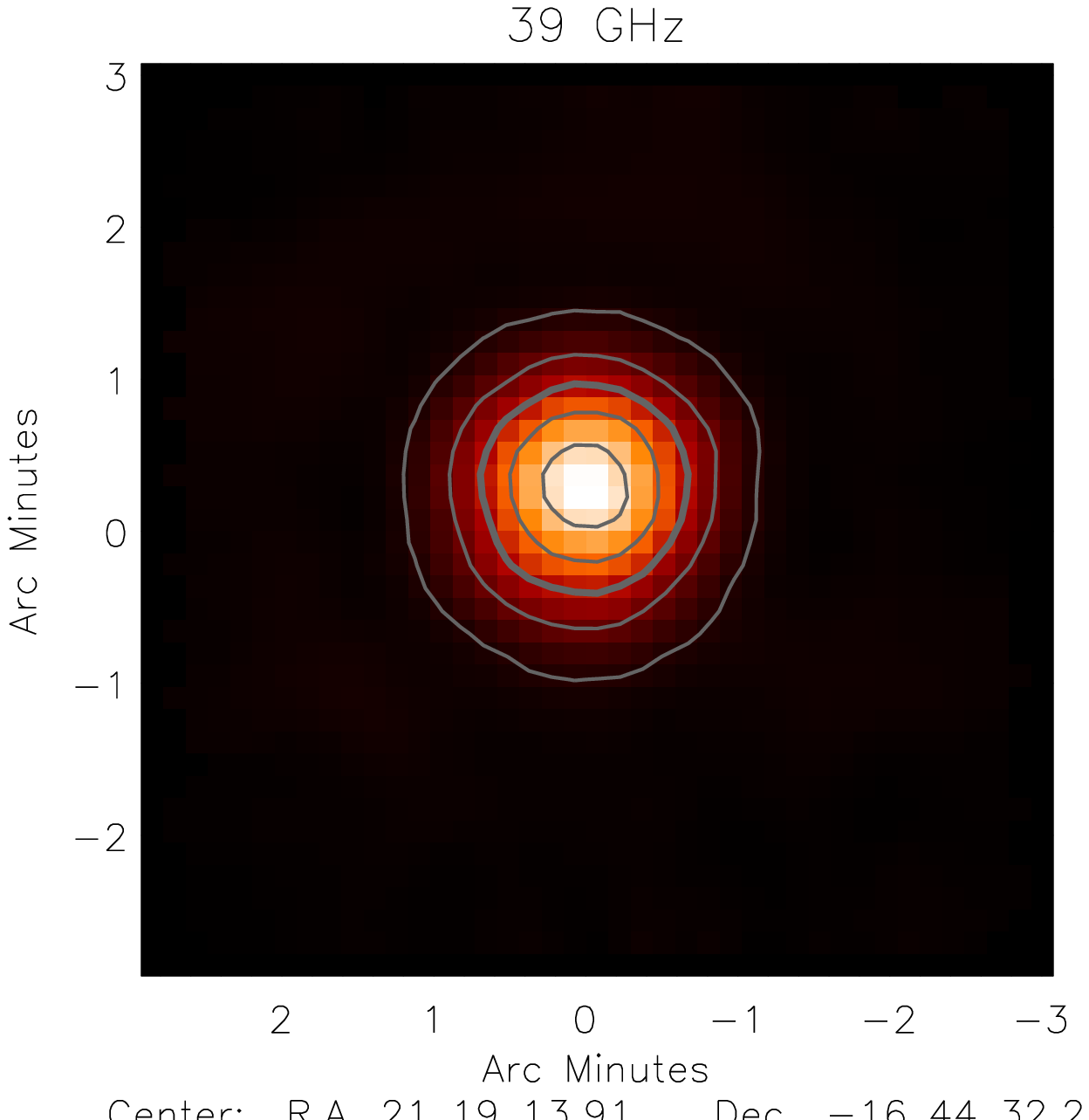}
\includegraphics[width=0.23\textwidth, trim=50 0 50 0]{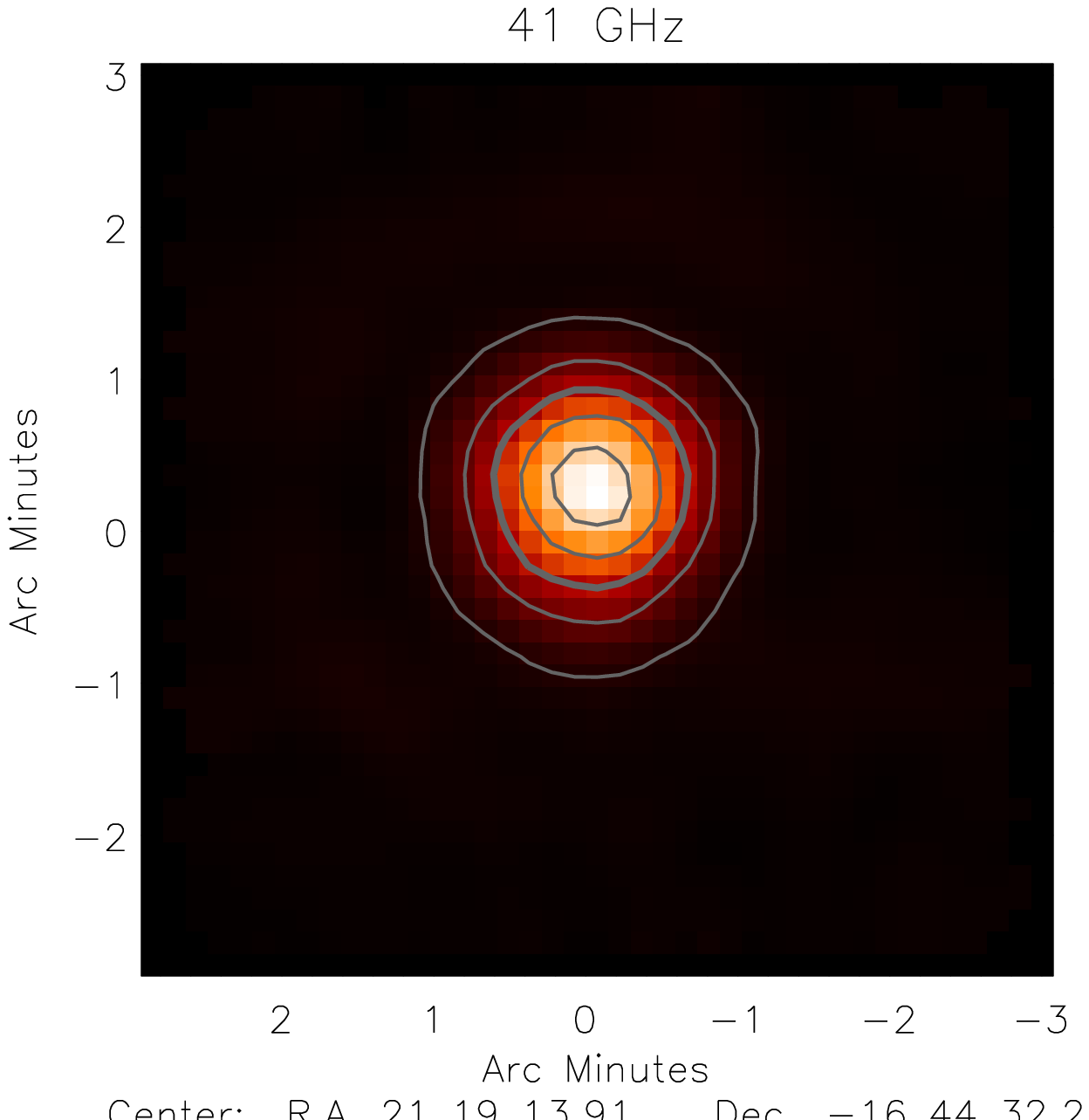}\\
\includegraphics[width=0.23\textwidth, trim=50 0 50 0]{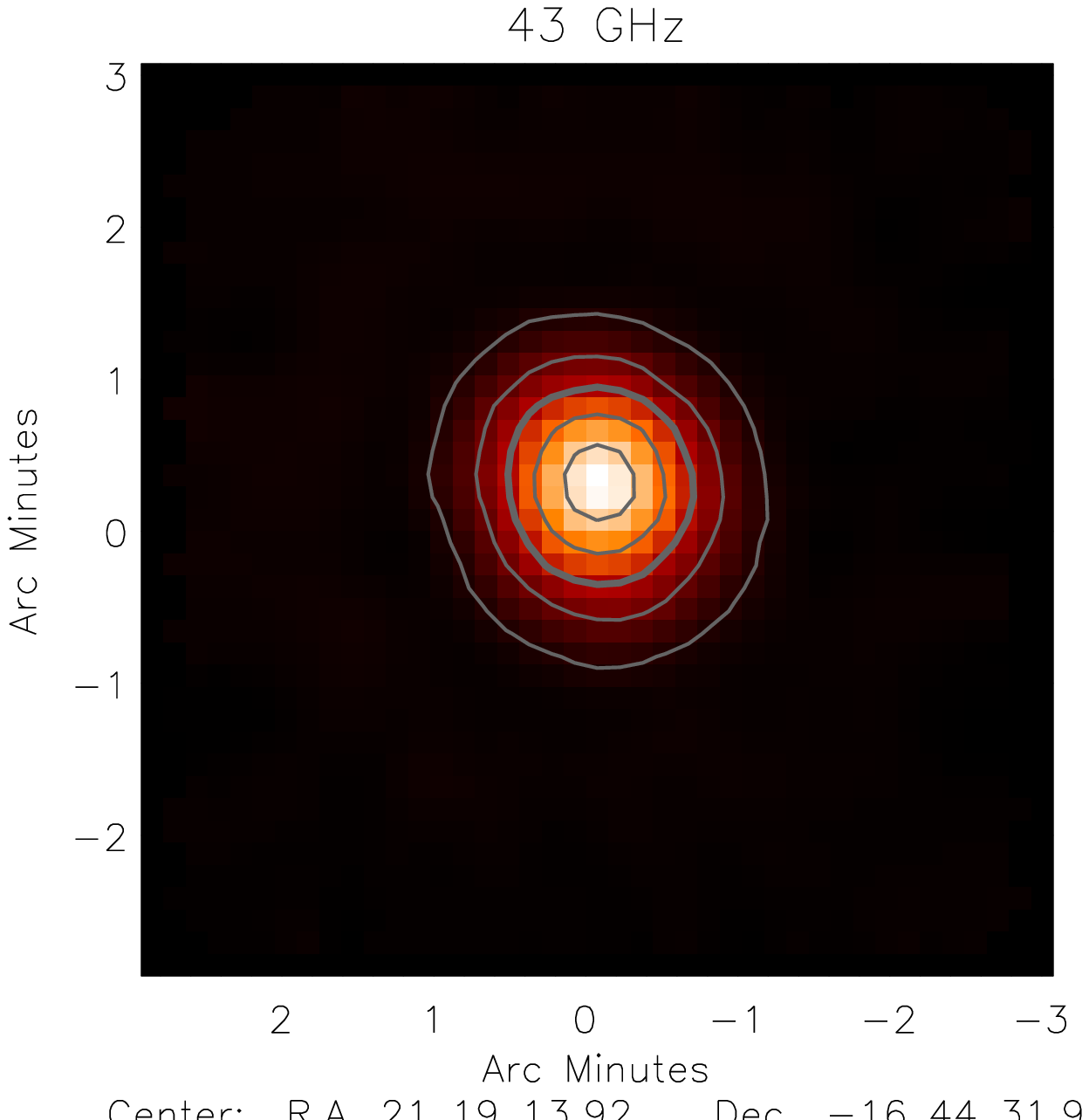}
\includegraphics[width=0.23\textwidth, trim=50 0 50 0]{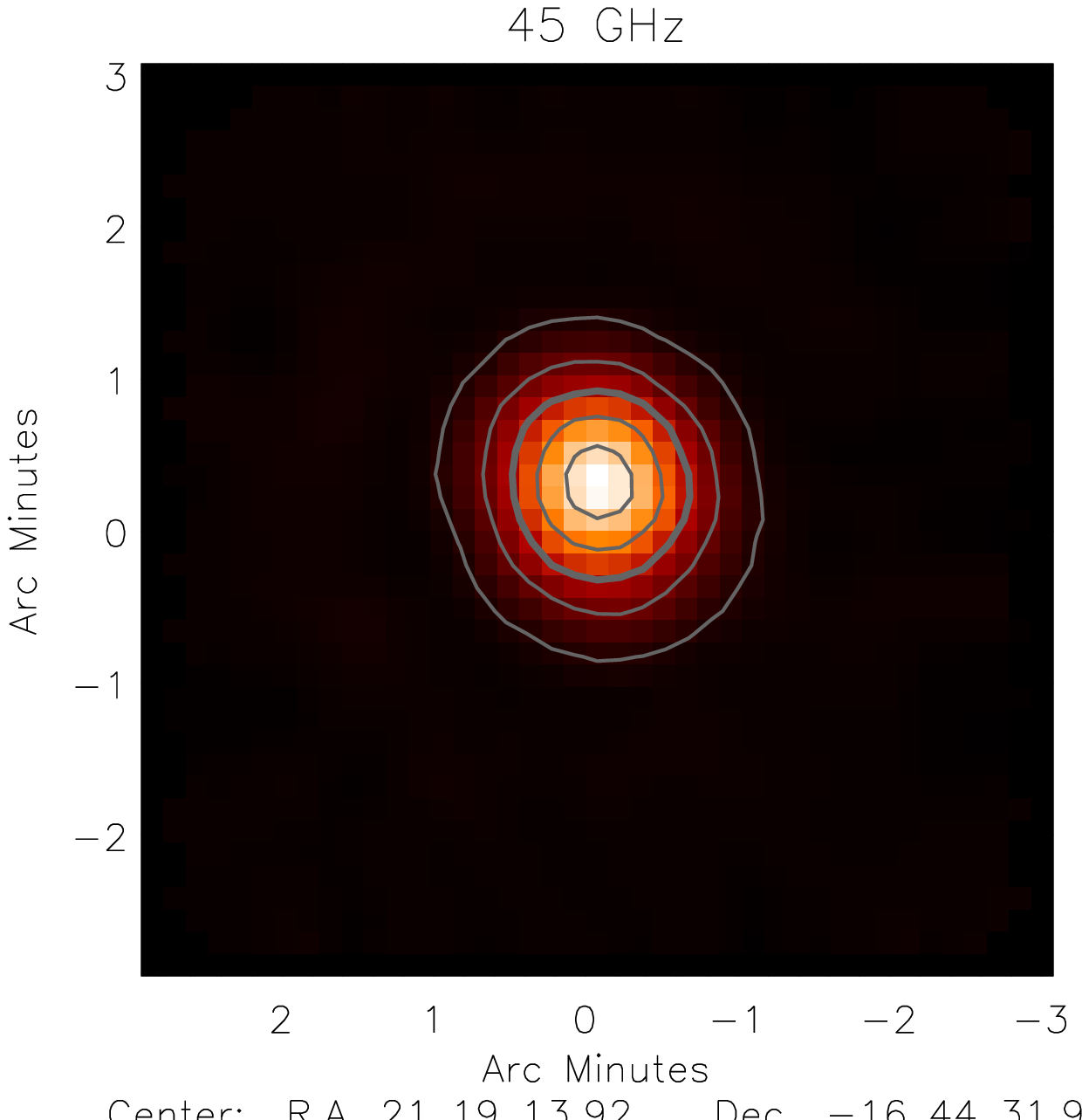}\\
\includegraphics[width=0.23\textwidth, trim=50 0 50 0]{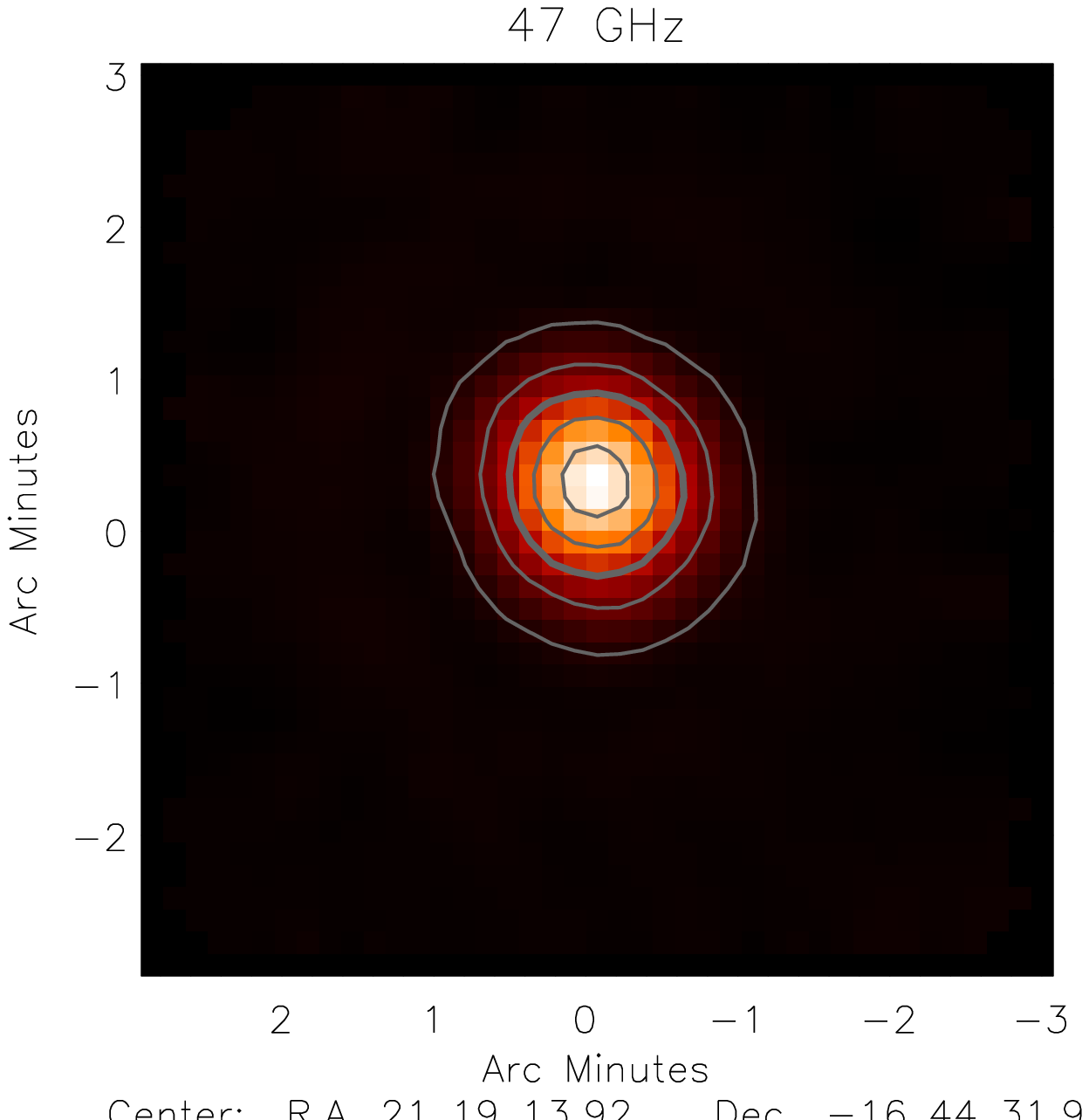}
\includegraphics[width=0.23\textwidth, trim=50 0 50 0]{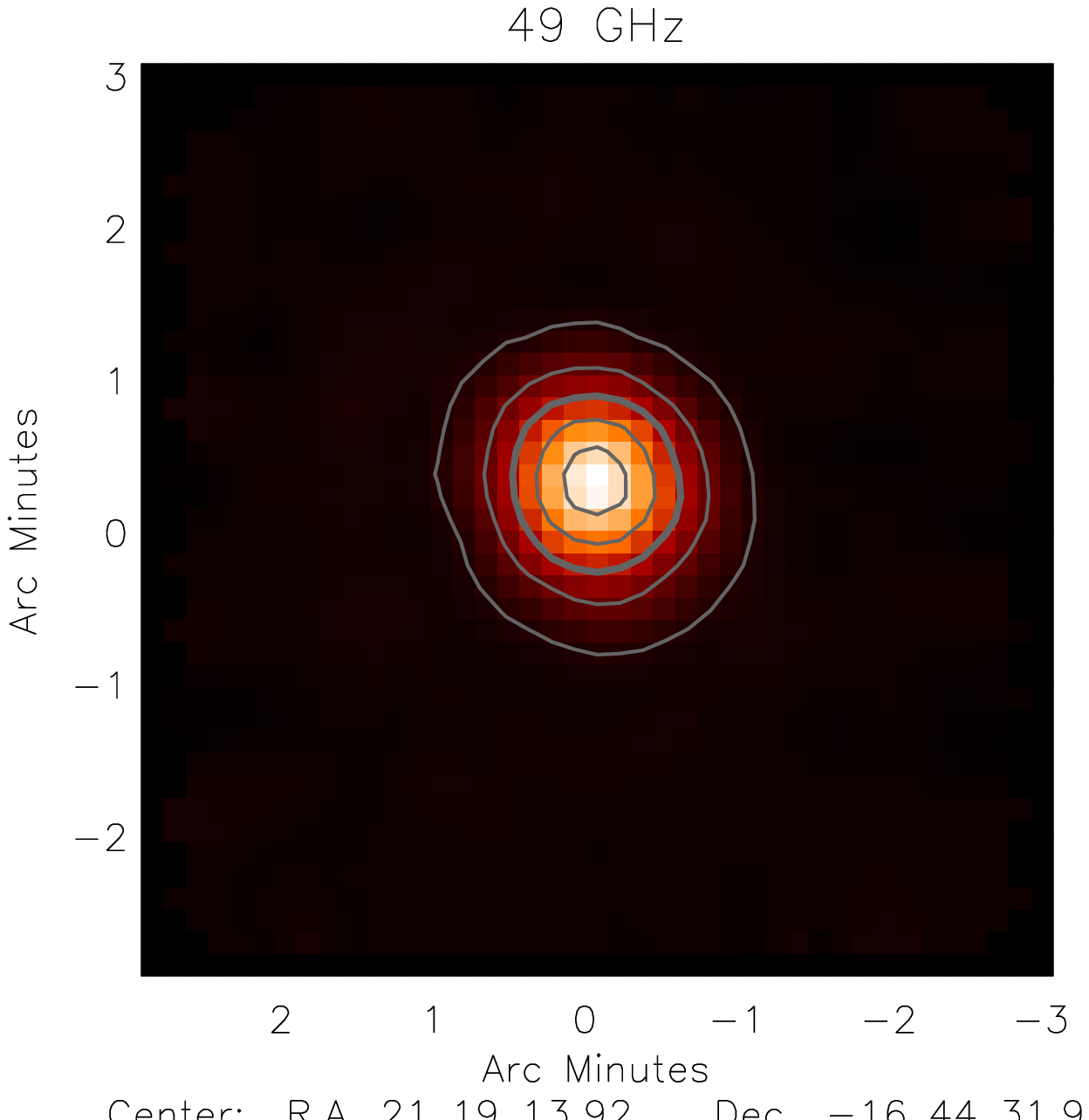}

\caption{Maps of the telescope beam produced from observations of Jupiter using the 7-mm receiver. Each map covers a 2\,GHz frequency window and are separated by 2\,GHz interval through the 30--50\,GHz frequency range of the 7-mm receiver. Contours have been added to illustrate the shape of the beam at the various frequencies; these start at 10\% of the peak emission and increase in steps of 20\%.}\label{fig:7mm_beam_maps}
\end{center}
\end{figure}

In Figures\,\ref{fig:12mm_beam_maps} and \ref{fig:7mm_beam_maps} we present the results of our beam mapping observations of Jupiter. The emission maps nicely illustrate the circular shape and the decrease of the telescope's FWHM beam size with increasing frequency, as expected. We note that in a number of the maps the position of the planet is offset from the pointing centre (0,0), which is the result of pointing errors. These are bigger than would normally be expected due to the large angular distance of the planet from the Galactic Plane during the observations, and consequently, the large angular distances between Jupiter and the nearest available pointing sources.  

As we have seen in the previous section the radial beam profile at both 12 and 7\,mm consists of three distinct components. The presence of these error beams somewhat complicates the calibration for sources with angular diameters larger than about twice the telescope's main beam. A consequence of this is that the telescope efficiency will depend on the angular size of the observed source. Following the lead of Ladd et al. (2005), we separate the calibration into two parts to accommodate both compact and extended sources.

\subsubsection{Main Beam Efficiency}

Planets are ideal astronomical sources to use for deriving beam efficiencies as their sizes and fluxes are well known. Moreover, given that Jupiter is unresolved at all frequencies of interest for this study it couples well with the telescope's main beam but not to either of the two error beams. In this cases the measured antenna temperature, $T^*_{\rm{A}}$, is a convolution of the sources brightness temperature distribution, $T_{\rm{B}}(\theta,\phi)$, with the telescope beam pattern, $P_{\rm{n}}(\theta,\phi)$, such that:   

\begin{equation}
T^*_{\rm{A}}=\frac{\int T_{\rm{B}}(\theta,\phi) P_{\rm{n}}(\theta,\phi)\,{\rm{d}}\Omega}{\int_{\rm{4\pi}} P_{\rm{n}}(\theta,\phi) \,{\rm{d}}\Omega}
\end{equation}

\begin{table*}[!ht]
\begin{center}
\caption{Summary of measured antenna temperatures, correction factors and beam efficiencies as a function of frequency. The columns are as follows: (1) observing frequency, (2) measured peak antenna temperatures, (3) and (4) beam dilution and coupling factors , (5) planet's diameter, (6) two numbers are presented, the first of which refers to the planets physical temperature taken from the model of de Pater \& Massie\,(1985), while the second number refers to the  Rayleigh-Jeans correction that needs to be applied to obtain the planets brightness temperature at a given frequency (i.e., $T_{\rm{B}}=T_{\rm{phys}}-\frac{h\nu}{2k}$), (7) expected antenna temperature for a prefect telescope, (8) conversion factor for antenna temperature and the flux density scale, (9) beam efficiency, $\eta_{\rm{mb}}$, calculated for compact sources which couple only to the telescopes main beam and  (10) the extended beam efficiency, $\eta_{\rm{xb}}$, for more extended sources ($>$ twice the FWHM of the main beam). The quoted errors are derived from the Gaussian fit to the azimuthal emission profile, however, taking into account systematic errors and atmospheric attenuation we estimate a more realistic error to be of order 10\%.}\label{tbl:efficiencies}
\begin{tabular}{cccccccccc}
\hline 
& 	 &  &  & \multicolumn{2}{c}{Planet Properties} & &Conversion\\
\cline{5-6}
Freq. 	& $T^*_{\rm{A}}$	 & $f$  & $K$ & D   & $T_{\rm{B}}$        & $T^{*\prime}_{\rm{A}}$&  Factor & $\eta_{\rm{mb}}$& $\eta_{\rm{xb}}$\\
(GHz)		& (K)				 &     &  &(\arcsec)  & (K)&    (K)     &(Jy K$^{-1}$)                   &                 \\
\hline
(1)	& (2)	 & (3)  & (4) & (5)   & (6)        & (7)&  (8) & (9)& (10) \\
\hline \hline

17	&	3.16$\pm$0.12	&	22.3	&	1.023	&	39.5	&	155$-$0.41	&	6.77	&	14.1	&	0.47$\pm$0.04	&	0.56$\pm$0.05	\\
19	&	3.85$\pm$0.14	&	18.9	&	1.027	&	39.5	&	146$-$0.46	&	7.50	&	12.9	&	0.51$\pm$0.04	&	0.62$\pm$0.04	\\
21	&	4.24$\pm$0.14	&	16.5	&	1.031	&	39.5	&	138$-$0.50	&	8.07	&	12.8	&	0.53$\pm$0.03	&	0.63$\pm$0.04	\\
23	&	5.00$\pm$0.16	&	14.0	&	1.036	&	39.5	&	134$-$0.55	&	9.17	&	12.3	&	0.54$\pm$0.03	&	0.65$\pm$0.04	\\
25	&	6.19$\pm$0.16	&	12.9	&	1.039	&	39.5	&	135$-$0.60	&	10.00	&	10.9	&	0.62$\pm$0.03	&	0.74$\pm$0.03	\\
27	&	7.77$\pm$0.18	&	10.3	&	1.049	&	39.5	&	138$-$0.65	&	12.74	&	11.1	&	0.61$\pm$0.02	&	0.73$\pm$0.03	\\
\hline
31	&	12.05$\pm$0.18	&	5.3	&	1.097	&	43.1	&	146$-$0.74	&	24.80	&	14.4	&	0.49$\pm$0.02	&	0.63$\pm$0.02	\\
33	&	14.55$\pm$0.15	&	4.7	&	1.110	&	43.1	&	149$-$0.79	&	28.32	&	13.7	&	0.51$\pm$0.01	&	0.67$\pm$0.01	\\
35	&	16.26$\pm$0.18	&	4.4	&	1.119	&	43.5	&	152$-$0.84	&	30.97	&	13.6	&	0.52$\pm$0.01	&	0.68$\pm$0.01	\\
37	&	17.87$\pm$0.15	&	4.0	&	1.129	&	43.5	&	154$-$0.89	&	33.60	&	13.6	&	0.53$\pm$0.01	&	0.69$\pm$0.01	\\
39	&	18.68$\pm$0.16	&	3.7	&	1.141	&	43.8	&	156$-$0.94	&	36.75	&	14.4	&	0.51$\pm$0.01	&	0.66$\pm$0.01	\\
41	&	21.21$\pm$0.17	&	3.3	&	1.161	&	43.8	&	158$-$0.98	&	41.41	&	14.5	&	0.51$\pm$0.01	&	0.67$\pm$0.01	\\
43	&	21.47$\pm$0.17	&	3.4	&	1.154	&	43.1	&	160$-$1.03	&	40.51	&	14.1	&	0.53$\pm$0.01	&	0.69$\pm$0.01	\\
45	&	22.27$\pm$0.17	&	3.1	&	1.167	&	43.1	&	161$-$1.08	&	43.52	&	14.8	&	0.51$\pm$0.01	&	0.67$\pm$0.01	\\
47	&	23.10$\pm$0.24	&	3.0	&	1.177	&	43.1	&	163$-$1.13	&	46.02	&	15.3	&	0.50$\pm$0.01	&	0.65$\pm$0.01	\\
49	&	21.61$\pm$0.19	&	2.7	&	1.196	&	43.1	&	164$-$1.18	&	50.27	&	18.1	&	0.43$\pm$0.01	&	0.56$\pm$0.01	\\
\hline
\end{tabular}
\end{center}
\end{table*}

For compact sources with a uniform brightness temperature the equation above simplifies to:

\begin{equation}
T^*_{\rm{A}} = \eta_{\rm{mb}} \frac{\int_{\rm{planet}} P_{\rm{n}}(\theta,\phi)\,{\rm{d}}\Omega}{\int_{\rm{mb}} P_{\rm{n}}(\theta,\phi)\,{\rm{d}}\Omega} T_{\rm{B}} = \eta_{\rm{mb}} \frac{\Omega_{\rm{planet}}} {\Omega_{\rm{mb}}} T_{\rm{B}}
\end{equation}

\noindent where $\Omega_{\rm{planet}}$ is the solid angle of the planet at the time of the observations, and $\Omega_{\rm{mb}}$ is the telescope main beam solid angle. $\eta_{\rm{mb}}$ is referred to as the main beam efficiency and is defined as the ratio of the total measured power within the telescope's main beam, and total measured power integrated over the full sphere of 4$\pi$, i.e.:

\begin{equation}
\eta_{\rm{mb}} = \frac{\int_{\rm{mb}} P_{\rm{n}}(\theta,\phi)\,{\rm{d}}\Omega}{\int_{4\pi} P_{\rm{n}}(\theta,\phi)\,{\rm{d}}\Omega} 
\end{equation}

The ratio of planetary and main beam solid angles shown in the right hand side of Equation\,3 consists of two important correction factors: beam dilution, $f$, which reduces the expected antenna temperature --- this can be large if the angular size of the main beam is much larger than angular size of the planet; and the coupling coefficient, $K$, which corrects for the antenna's response for the non-pointlike planetary disk. Beam dilution is simply the ratio of the area of the planetary disk ($\pi/4\times D^2_{\rm{planet}}$ --- where $D$ is the angular diameter of the planet) and the telescope's main beam ($1.133\times \theta^2_{\rm{mb}}$): 

\begin{equation}
f^{-1}=\frac{D^2_{\rm{planet}}} {\theta^2_{\rm{mb}}}{\rm{ln}} 2 
\end{equation}

\noindent and $K$ can be calculated following Baars (1973) Equation\,12:

\begin{equation}
K = \frac{x^2}{1-{\rm{exp}}(-x^2)}\,{\rm{where}}\,x=\frac{D_{\rm{planet}}}{\theta_{\rm{mb}}}\sqrt{{\rm{ln}}2}
\end{equation}

Both correction factors $K$ and $f$ have been derived using the angular diameter of the planet at the time of the observation, and the  FWHM of the telescopes main beam has been derived from a Gaussian fit to the azimuthal radial profiles of the Jupiter maps (i.e., using Equation\,1). The derived values of $K$ and $f$ and the angular size of the planet at the time of the various observations are presented in Columns\,3, 4 and 5 of Table\,3, respectively. 

\begin{figure*}
\begin{center}
\includegraphics[width=0.95\textwidth]{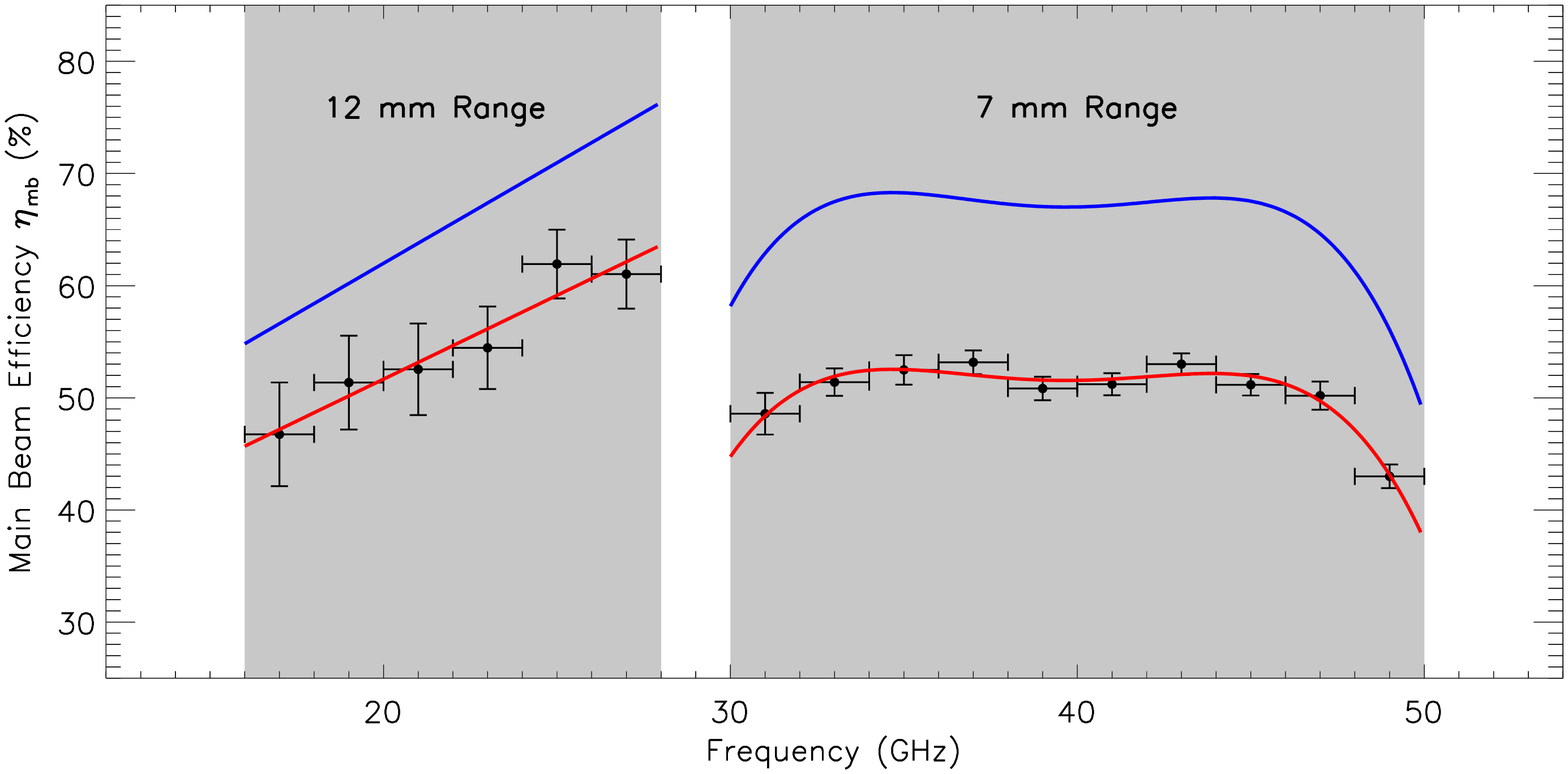}

\caption{Measurements of the Mopra telescope beam efficiencies for the 12 and 7-mm receivers. The vertical error bars result from the propagation of the 1$\sigma$ error associated with the measured peak antenna temperature, while the horizontal error bars simply indicate the frequency bandpass over which the measurement is applicable.  The regions shaded in grey indicate the frequency range covered by the 12- and 7-mm receivers and are labelled accordingly. The results of a polynomial fit to the measured main beam efficiency ($\eta_{\rm{mb}}$) for each frequency range are plotted in red. The blue line indicates the estimated efficiency of the extended beam (see Section 3.3.2 for details).}\label{fig:efficiency_plot}
\end{center}
\end{figure*}

\begin{sloppypar}
The frequency-dependent brightness temperature of Jupiter is derived from the model of de Pater \& Massie\,(1985), 
which gives the physical temperature, $T_{\rm{phys}}$.  The planet's brightness temperature is 
obtained by applying the Rayleigh-Jeans correction: $T_{\rm{B}}=T_{\rm{phys}}-\frac{h\nu}{2k}$ --- the planets physical temperature and Rayleigh-Jeans correction are given in this form in Column\,6 in Table\,3. The correction is $\sim$0.5 and 1\,K for the 12- and 7-mm receiver ranges respectively. This model has been 
adopted because it yields brightness temperatures at all frequencies and is also used for calibration at the 
Australia Telescope Compact Array. The model is quite consistent with recent broadband measurements of the 
brightness temperature of Jupiter: 152$\pm$5\,K at 32\,GHz (6\,GHz bandwidth, Mason et al.\,1999); 134$\pm$4,
146.6$\pm$2.0, and 154.7$\pm$1.7\,K at 23, 33, and 41\,GHz, respectively (Page et al.\ 2003).
Jupiter's brightness temperature, the beam dilution and coupling coefficient together yield the expected antenna 
temperature, ${T^{*\prime}_{\rm{A}}}$, assuming a perfect dish and loss-less front and backends, i.e.;
\end{sloppypar}

\begin{equation}
T^{*\prime}_{\rm{A}} = \frac{T_{\rm{B}}}{fK}
\label{eqn:efficiency}
\end{equation}

\begin{sloppypar}
The expected antenna temperatures calculated using the equation above are presented in Column\,7 of Table\,3.  In Column\,8 of Table\,3 we present the factor required to convert between antenna temperature and the flux density scale for point source estimated from our Jupiter measurements. Finally, we estimate the main beam efficiency using:
\end{sloppypar}

\begin{equation}
\eta_{\rm{mb}}= \frac{T^*_{\rm{A}}}{T^{*\prime}_{\rm{A}}}
\label{eqn:efficiency}
\end{equation}

\noindent The main beam efficiencies can be found in Column\,9 of Table\,3. 

In Figure\,\ref{fig:efficiency_plot} we present a plot of the main beam efficiency as a function of frequency. The individual data points are shown as filled circles and the frequency ranges for each of the receivers are indicated by grey shading. The main beam efficiencies obtained for the 12-mm receiver show an increasing efficiency with frequency beginning at slightly below 50\% at 16\,GHz rising to $\sim$65\% at the upper frequency range of 27\,GHz. The 12-mm data can be reasonably well fit with a straight line $y=A+Bx$ where $A=21.76$ and $B=1.49$ and $x$ is the observed frequency in GHz. 

The profile of the beam efficiencies over the 7-mm range is a little more complicated. It appears reasonably flat between 33 and 47\,GHz with an average main beam efficiency of $\sim$50\%, with the efficiency falling off towards either end of the frequency band. The 7-mm data is best fit with a fourth order polynomial, $y=A+Bx+Cx^2+Dx^3+Ex^4$ where $A=-4085.61$,  $B=425.97$, $C=-16.3656$, $D=0.27813$ and $E=-0.00176445$ and $x$ is as previously defined.

\subsubsection{Mopra Extended Beam Efficiency}

As we have seen in Section\,3.2 the telescope beam pattern consists of three components we referred to as the main beam, and inner and outer error beams. Compact sources couple to the main beam only. However, extended sources (larger than $\sim$2.8\arcmin\ and 1.6\arcmin\ at 12- and 7-mm respectively) can couple to the inner and outer error beams as well, making the calibration more complicated. Following the method of Ladd et al.\,(2005) we write an expression for the measured antenna temperature for extended sources as a function of the extended beam pattern, $P_{\rm{xb}}$, which included contribution from both the main beam and error beams such that:

\begin{equation}
T^*_{\rm{A}} = \eta_{\rm{xb}} \frac{\int T_{\rm{B}}(\theta,\phi) P_{\rm{xb}}(\theta,\phi)\,{\rm{d}}\Omega}{\int_{\rm{mb}} P_{\rm{xb}}(\theta,\phi)\,{\rm{d}}\Omega}  
\end{equation}

\noindent where  $\eta_{\rm{xb}}$ is the extended beam efficiency defined as:

\begin{equation}
\eta_{\rm{xb}} = \frac{\int_{\rm{xb}} P_{\rm{xb}}(\theta,\phi)\,{\rm{d}}\Omega}{\int_{4\pi} P_{\rm{n}}(\theta,\phi)\,{\rm{d}}\Omega} 
\end{equation}

Similarly to Ladd et al. (2005) we have no direct measurements of extended sources, however, we can use the radial beam patterns obtained from the maser beam maps presented in Figure\,3 to estimate the extended beam efficiency. Referring to Table\,2, we estimate the amount of power within the inner and outer error beams is approximately 20\% and 24\% of that found in the main beam at 22 and 43\,GHz, respectively. These values would be reduced to 15\% at both frequencies if the source couples to the main beam and the inner error beam only. Using our knowledge of the radial beam pattern we estimate the fraction of the total power contained within the extended beam is $\sim$60--75\% for the 12-mm receiver range and $\sim$65\% for the 7-mm receiver. We present the values of the extended beam efficiency in the final column of Table\,3 and are also shown on Figure\,6 (blue line).

\section{Summary}

We present the results of a set of mapping and pointed observations towards astronomical masers and Jupiter chosen to characterise the performance of the Mopra Radio Telescope at 12 and 7\,mm. We calculate the telescope's main beam size and efficiencies over the  16--50\,GHz frequency range. We find that the beam size is well fit by $\lambda$/$D$ over the frequency range with a correlation coefficient of $\sim$90\%. Telescope main beam efficiencies for the 12-mm receiver range between $\sim$48--64\%, with the 7-mm receiver having a beam efficiency of $\sim$50\% over the majority of the band. 

Using beam maps of strong H$_2$O and SiO masers we investigate the  telescope's radial beam pattern. At both frequencies the radial beam pattern reveals the presence of three components, a central `core', which is well fit by a Gaussian and constitutes the telescope main beam, and inner and outer error beams. At both frequencies the inner and outer error beams extend out to approximately 2 and 3.4 times the full-width half maximum of the main beam respectively. From measurements of the radial beam power pattern we estimate the amount of power contained in the inner and outer error beams is 20\% at 22\,GHz rising to 30\% at 43\,GHz. Sources with angular sizes a factor of two or more larger than the telescopes main beam will couple to the main and error beams, and thus, the power contributed by the error beams needs to be considered. We therefore also provide an estimate of the extended beam efficiencies. 

\begin{sloppypar}
Since the characterisation of this telescope is an ongoing process it is conceivable that some of these results may change in future due to adjustments of the focus or subreflector. Moreover, as mentioned earlier, due to limited observing time dependence of the results presented here have not been fully tested as a function of elevation or azimuth. We have therefore set up the following web page which will be periodically updated as new data becomes available and our calibration is refined: www.narrabri.atnf.csiro.au/mopra/calibration/mopra-calib.html.
\end{sloppypar}

\section*{Acknowledgments} 

The authors would like to thank the staff of the Paul Wild Observatory for their assistance during the preparation of our observations. We are grateful to Mike Kesteven for technical help with the mapping observations. The would also like to thank the annonymous referee for some useful comments and suggestions. JSU is supported by a CSIRO OCE fellowship. The Mopra Telescope is part of the Australia Telescope and is funded by the Commonwealth of Australia for operation as a National Facility managed by CSIRO. The University of New South Wales Mopra Spectrometer Digital Filter Bank used for the observations with the Mopra Telescope was provided with support from the Australian Research Council, together with the University of New South Wales,
University of Sydney, Monash University and the CSIRO.

\end{document}